\documentclass[preprint]{aastex}

\usepackage{graphicx}
\usepackage{dcolumn}
\usepackage{bm}
\usepackage{amsfonts}
\usepackage{color}
\usepackage{amsmath}

\begin{document}

\shorttitle{Correlated Signatures of GW and Neutrino Emission in 3D-GR CCSN 
 Simulations}
\shortauthors{Kuroda, Kotake, Hayama, \& Takiwaki}

\title{Correlated Signatures of Gravitational-Wave and Neutrino Emission in Three-Dimensional General-Relativistic Core-Collapse Supernova Simulations}

\author{Takami Kuroda\altaffilmark{1}, Kei Kotake\altaffilmark{2,3}, 
Kazuhiro Hayama\altaffilmark{4}, and 
Tomoya Takiwaki\altaffilmark{5}}

\affil{$^1$Institut f{\"u}r Kernphysik, Technische Universit{\"a}t Darmstadt,
Schlossgartenstrasse 9, D-64289 Darmstadt, Germany}
\affil{$^2$Department of Applied Physics, Fukuoka University,
 8-19-1, Jonan, Nanakuma, Fukuoka, 814-0180, Japan}
\affil{$^3$Max Planck Institut f\"ur Astrophysik, Karl-Schwarzschild-Str. 1, D-85748, Garching, Germany}
\affil{$^4$KAGRA Observatory, Institute for Cosmic Ray Research, University of Tokyo, 238 Higashi Mozumi, Kamioka, Hida, Gifu 506-1205, Japan} \affil{$^5$Division of Theoretical Astronomy, National Astronomical Observatory of Japan, 2-21-1, Osawa, Mitaka, Tokyo, 181-8588, Japan}

\begin{abstract}
We present results from general-relativistic (GR) three-dimensional (3D) core-collapse
 simulations with approximate neutrino transport for three non-rotating progenitors (11.2, 15, and 40 $M_{\odot}$) using different nuclear equations of state (EOSs). 
We find that 
the combination of progenitor's higher compactness at bounce and the use of 
  softer EOS leads to stronger activity of the standing accretion shock 
instability (SASI). We confirm previous predications that the SASI produces 
 characteristic time modulations 
both in neutrino and gravitational-wave (GW) signals. By performing 
 a correlation 
 analysis of the SASI-modulated neutrino and GW signals, we find that the 
correlation becomes highest when we take into account the 
  time-delay effect due to the advection of material from 
the neutrino sphere to the proto-neutron star core surface. Our results suggest that 
the correlation of the neutrino and GW signals, if detected, would provide a 
 new signature of the vigorous SASI activity in the supernova core,
 which can be hardly seen if neutrino-convection dominates over the SASI.
\end{abstract}


\section{Introduction}
Core-collapse supernovae (CCSNe) have been attracting attention
 of theoretical and observational astrophysicists for many decades. 
From multi-wavelength electromagnetic (EM) wave signals, a wide
 variety of observational evidence have been reported so far including
 ejecta/line morphologies, spatial distributions 
of heavy elements, and proper motions of pulsars, which
 have all pointed toward CCSNe being generally aspherical (e.g., 
\citet{larsson16,grefenstette2017,tanaka17,lopez17} and references therein). 
Unambiguously important these discoveries are, the EM signals 
could only provide an indirect probe of 
the explosion mechanism of CCSNe, because they 
 snapshot images of optically thin regions far away from the central engine.

 Neutrinos and gravitational waves (GWs) are expected to provide
 direct probes of the inner-workings of CCSNe (e.g., \citet{mirizzi16,Kotake13}
 for a review). Currently multiple neutrino detectors capable 
of detecting CCSN neutrinos are in operation
(e.g., \citet{scholberg12} for a review). The best suited detectors are 
{\it Super-Kamiokande} (Super-K) and IceCube that can detect rich dataset
 of neutrino events (for example $\sim 10^4$ for Super-K) from future Galactic CCSNe 
 \citep{Ikeda07,IceCube11}. In the past thirty years after SN1987A - the only 
CCSN with neutrino detection to date \citep{hirata87,bionta87}, 
significant progress has been 
 also made in GW detectors (e.g., \citet{schutz09} for a review).
 The sensitivity has been significantly enhanced enough to 
 allow the first detection by the LIGO 
collaboration for the black hole merger event \citep{gw2016}. 
The second-generation detectors like advanced VIRGO
\citep{advv} and
 KAGRA \citep{Aso13} will be online in the coming years.
 Furthermore third-generation detectors like 
 Einstein Telescope and Cosmic Explorer are recently being proposed 
\citep{punturo,CE}. At such a high level of sensitivity, CCSNe are also 
expected as one of the most promising astrophysical sources 
of GWs (e.g., \cite{kotake_kuroda16,fryer11,Ott09} for review).

From a theoretical point of view, neutrino radiation-hydrodynamics simulations of CCSNe
are converging to a point that multi-dimensional (multi-D)
 hydrodynamics instabilities including neutrino-driven convection 
(e.g., \citet{couch13a,Murphy13,Hanke12}) and the 
Standing-Accretion-Shock-Instability (SASI, \citet{Blondin03,Foglizzo06,Fernandez15})
 play a crucial key role in facilitating the 
neutrino mechanism of CCSNe \citep{bethe}.
In fact, a number of self-consistent 
 models in two or three spatial dimensions (2D, 3D)
 now report revival of the stalled bounce shock into explosion
 by the ``multi-D'' neutrino mechanism (see \cite{janka17,bernhard16,thierry15,burrows13,Kotake12} for reviews)\footnote{Here we shall consider canonical CCSN progenitors \citep{Heger05} 
where rotation/magnetic fields play little role in the explosion dynamics 
(see, however, \citet{moesta14,takiwaki16,martin17}).}. 

Conventionally the GW and neutrino signatures 
from CCSNe have been studied rather separately. 
 For the neutrino signals, 
\citet{Tamborra13} were the first to find the SASI-induced modulations in the 
 neutrino signals using results from full-scale 3D CCSN models \citep{Hanke13}. 
They found that the SASI-induced modulation is clearly visible 
 for two high-mass progenitors ($20$ and $27 M_{\odot}$) where high SASI activity 
 was observed in the postbounce (pb) phase \citep{Tamborra14}. They pointed out that 
the frequency of the SASI-induced neutrino emission peaks around 
$\sim80$ Hz, which can be detectable by IceCube or the future 
{\it Hyper-Kamiokande} (Hyper-K) for a Galactic event at a distance of $\sim 10$ kpc 
(see also \citet{marek09gw,lund10,brandt,bernhard14a}). 

From recent self-consistent 3D models, it becomes clear that the 
SASI also produces a characteristic signature in the GW emission
 \citep{KurodaT16ApJL,Andresen17}. There are several GW emission processes 
in the postbounce phase including prompt convection, 
neutrino-driven convection, proto-neutron star (PNS) convection, the SASI, and $g$-mode
 oscillation of the PNS surface 
(e.g., \citet{EMuller97,EMuller04,Murphy09,Kotake09,BMuller13,CerdaDuran13}).
 Among them, the most distinct GW emission process generically seen in recent 
self-consistent CCSN models 
 is the one from the PNS surface oscillation \citep{BMuller13,KurodaT16ApJL,Andresen17,Yakunin17}.
 The characteristic GW frequency increases almost monotonically 
with time due to an accumulating accretion to the PNS, which ranges 
from $\sim 100$ Hz to $\sim 1000$ Hz in the typical simulation timescales. 
On the other hand, 
the GW frequency from the SASI appears in the lower frequency range of $\sim 100$ to $250$ Hz 
and persists when the SASI dominates over neutrino-driven convection \citep{KurodaT16ApJL,Andresen17}.
 \citet{Andresen17} pointed out that third-generation detectors (like ET) could 
distinguish SASI- from convection-dominated case among their full-scale 3D models 
\citep{Hanke13,Melson15b} at a distance of $\sim$ 10 kpc.

These findings may raise a simple question whether there is some correlation between 
the SASI-induced neutrino and GW signals. Spotted by the neutrino and GW astronomy in the 
advanced era, the time is ripe to study in detail what we can learn about the explosion mechanism 
from the future {\it simulatenous} detection of neutrinos and GWs using outcomes 
of multi-D CCSN models. In our previous work \citep{KurodaT16ApJL}, 
we have investigated the GW signatures based on 3D full-GR simulations with 
 approximate neutrino transport for a non-rotating $15 M_{\odot}$ star, 
using three different EOSs.
 In this work, we will compute two more progenitors of low- or high- progenitor compactness (11.2 and 40 $M_{\odot}$). Following \citet{Tamborra14}, we estimate 
 neutrino event rates in both Hyper-K and IceCube from our 3D-GR models. 
We perform a correlation analysis between the GW and neutrino signals. 
We discuss what we can learn about the
 supernova engine if the simultaneous detection is made possible for a next 
 CCSN event.

 This paper is organized as follows. We first give 
 a short summary of the numerical setup and the extraction of GWs in Section \ref{sec2}.
In Section \ref{sec3}, we present a short overview of hydrodynamics of our models. We then present analysis on the GW signatures in Section \ref{sec4}.
  The correlation 
 analysis between the GW and neutrino signals is presented in Section \ref{sec5}. We summarize the results
 and discuss its implications in Section \ref{sec6}.

\section{Numerical Methods and Initial Models}
\label{sec2}
The numerical schemes for our 3D-GR models are essentially the same as those 
in \citet{KurodaT16ApJL}.
For the metric evolution, we employ the standard BSSN variables 
($\tilde\gamma_{ij}$, $\phi$, $\tilde A_{ij}$, $K$ and $\tilde\Gamma^{i}$ \citep{Shibata95,Baumgarte99}).
Solving the evolution equations of metric, hydrodynamics, 
and neutrino radiation in an operator-splitting manner, the system evolves self-consistently
as a whole satisfying the Hamiltonian and momentum constraints \citep{KurodaT12,KurodaT14}. 
The total stress-energy tensor is $T_{\rm (total)}^{\alpha\beta} = T_{\rm (fluid)}^{\alpha\beta} + 
 \sum_{\nu}T_{(\nu)}^{\alpha\beta}$, where $T_{\rm (fluid)}^{\alpha\beta}$ and $T_{(\nu)}^{\alpha\beta}$ are the 
stress-energy tensor of fluid and the neutrino radiation field, respectively.
We consider three flavors of neutrinos ($\nu\in\nu_e,\bar\nu_e,\nu_x$) with $\nu_x$ 
representing heavy-lepton neutrinos (i.e. $\nu_{\mu}, \nu_{\tau}$ and their anti-particles).
All radiation and hydrodynamical variables are evolved in a conservative form.
To follow the 3D hydrodynamics up to $\lesssim 200$ ms postbounce,\footnote{Note in \citet{KurodaT16ApJL} the data up to 350 ms postbounce were
 shown. However, we are only able to compute up to $\sim$ 200 ms 
postbounce for the newly added models ($11.2$ and $40 M_{\odot}$)
 simply due to limited computational resources.
 To make the comparison fair (especially regarding the detectability 
(Figure \ref{f4})), we shall often limit the analysis 
up to $\sim$ 200 ms postbounce in this work (see, however, Figure \ref{f6}).}
 we shall omit the energy dependence of the 
 radiation in this work (see, however, \cite{KurodaT16,roberts16}).

We use three EOSs based on the 
relativistic-mean-field theory with different
 nuclear interaction treatments, which are DD2 and TM1 of \cite{HS} 
and SFHx of \citep{SFH}. For SFHx, DD2, and TM1\footnote{The symmetry energy $S$ at nuclear saturation density
is $S=28.67$, 31.67, and 36.95 MeV, respectively. \citep[e.g.,][]{Fischer14}},
the maximum gravitational mass ($M_{\rm max}$) and
the radius ($R$) of cold neutron star (NS) in the vertical part of the mass-radius relationship are
 $M_{\rm max}=2.13$, 2.42, and, 2.21 $M_\odot$
and $R\sim12$, 13, and, 14.5 km, respectively \citep{Fischer14}.
SFHx is thus softest followed in order by DD2, and TM1.
Among the three EOSs, DD2 is constructed in a way that fits well with 
  nuclear experiments \citep{Lattimer13},
 whereas SFHx is the best fit model with the observational mass-radius relationship 
\citep{SFH}.
All EOSs are compatible with the $\sim 2 M_{\odot} $ NS mass measurement \citep{Demorest10,Antoniadis}.

We study frequently used solar-metallicity models of a 15 $M_{\odot}$ star \citep{WW95},
 an 11.2 $M_{\odot}$ and a 40 $M_{\odot}$ star of \citet{WHW02}, respectively.
The 3D computational domain is a cubic box with 15000 km width
and nested boxes with 8 refinement levels are embedded.
Each box contains $128^3$ cells and the minimum grid size near the origin is $\Delta x=458$m.
In the vicinity of the stalled shock at a radius of $\sim100$ km, our resolution 
achieves $\Delta x\sim 1.9$ km, i.e., the effective angular resolution becomes $\sim1^\circ$.
Our 3D-GR models are named by the progenitor mass with the EOS in parenthesis like
S15.0(SFHx) which represents the progenitor mass of 15.0 $M_\odot$  and the EOS SFHx are used.

We extract GWs from our simulations using the conventional quadrupole formula
 \citep{Misner73}. The transverse and the trace-free gravitational field $h_{ij}$ is, 
\begin{eqnarray}
\label{eq:hij}
h_{ij}(\theta,\phi)=\frac{A_+(\theta,\phi)e_++A_\times(\theta,\phi) e_\times}{D},
\end{eqnarray}
where $A_{+/\times}(\theta,\phi)$ represent amplitude of orthogonally polarized wave 
components with emission angle $(\theta,\phi)$ dependence
 \citep{EMuller97,Scheidegger10,KurodaT14}, $e_{+/\times}$ denote unit polarization
 tensors. In this work, we extract GWs along the north pole 
$(\theta,\phi)=(0,0)$ and assume a source at a distance of 
 $D=10$ kpc.


\section{Overview of Hydrodynamics Features}
\label{Sec:Overview of Hydrodynamics Features}
In this section, we first present a short overview of hydrodynamics features 
in our 3D models for later convenience.

\label{sec3}
\begin{table}[htpb]
\begin{center}
\begin{tabular}{c cccc}
\hline\hline
Model&$\xi_{1.5}$ & $\rho_{\rm{c,cb}}(10^{14}$ g cm$^{-3}$) & $M_{\rm cb}(M_\odot)$&$M_{\rm cb}/R_{\rm cb}$(\%)\\
\hline
S15.0(SFHx)&0.592 &  4.50 &0.751& 7.72\\
S15.0(DD2)&0.592 &  3.75 &0.749& 5.21\\
S15.0(TM1)&0.592 &  3.69 &0.688& 4.51\\
S11.2(SFHx)&0.195 &  4.23 &0.663& 4.84\\
S40.0(SFHx)&0.990 &  4.47 &0.765& 5.07\\
\hline
\end{tabular}
\caption{Progenitor's compactness parameter ($\xi_{1.5}$)
 and key quantities at core bounce (labeled as ``cb'' in the table) 
for all the computed models. Except for the compactness parameter (see text for 
 definition), 
the maximum (rest-mass) density $\rho_{\rm{c,cb}}$,
the (unshocked) core mass $M_{\rm cb}$, and its non-dimensional relativistic parameter $M_{\rm cb}/R_{\rm cb}(=GM_{\rm cb}/c^2R_{\rm cb}$ in the cgs unit) are estimated at core bounce.}
\label{tb:BounceProfile}
\end{center}
\end{table}

Table \ref{tb:BounceProfile} compares progenitor's compactness parameter 
\citep{oconnor,Nakamura15} and several key quantities at core bounce 
for all the computed models in this work. For the compactness parameter, 
we adopt $M_{\rm bary}=1.5M_{\odot}$ at $t=0$ of 
 $\xi_{1.5} \equiv M_{\rm bary}/M_{\odot}(R(M_{\rm bary}=M)/1000{\rm km})^{-1}$ 
in the table. For the given progenitor mass (S15.0), one can see that the 
maximum density $\rho_{\rm c,cb}$ becomes higher for model with softest EOS (SFHx).
This is consistent with \citet{Fischer14}. 
Also the compactness parameter at bounce ($M_{\rm cb}/R_{\rm cb}$) has a 
correlation with the stiffness of the EOS. This is because the softer EOS leads 
to more compact and massive unshocked core, which makes $M_{\rm cb}/R_{\rm cb}$ higher.
 For the given EOS (SFHx), one can also see that the initial core 
compactness ($\xi_{1.5}$) has non monotonic impact on the compactness 
at bounce (compare $M_{\rm cb}/R_{\rm cb}$ for S15.0(SFHx) and S40.0 
(SFHx)). This is simply because the higher density profile 
in the precollapse phase leads to more massive inner-core 
(compare $\xi_{1.5}$ with $M_{\rm cb}$ in the table),
 which also makes the radius of the forming bounce shock bigger.

\begin{figure}[htpb]
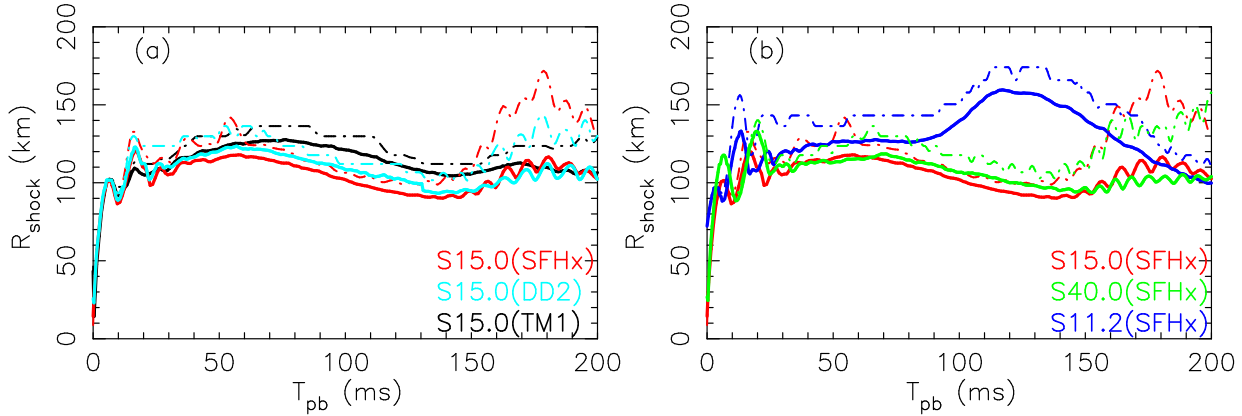

\begin{center}
\includegraphics[width=55mm,angle=-90.]{f6.eps}
\includegraphics[width=55mm,angle=-90.]{f7.eps}
  \caption{Time evolution of average (solid line) and maximum 
(dotted-dashed line) shock radii for all the models. 
The left and right panel compares the effect of EOSs and the 
progenitor masses, respectively.
  \label{f1}}
\end{center}
\end{figure}

\begin{figure*}[htbp]
  \begin{center}
    \begin{tabular}{cc}
      \begin{minipage}[t]{0.5\hsize}
        \begin{center}
          \includegraphics[clip,width=70mm]{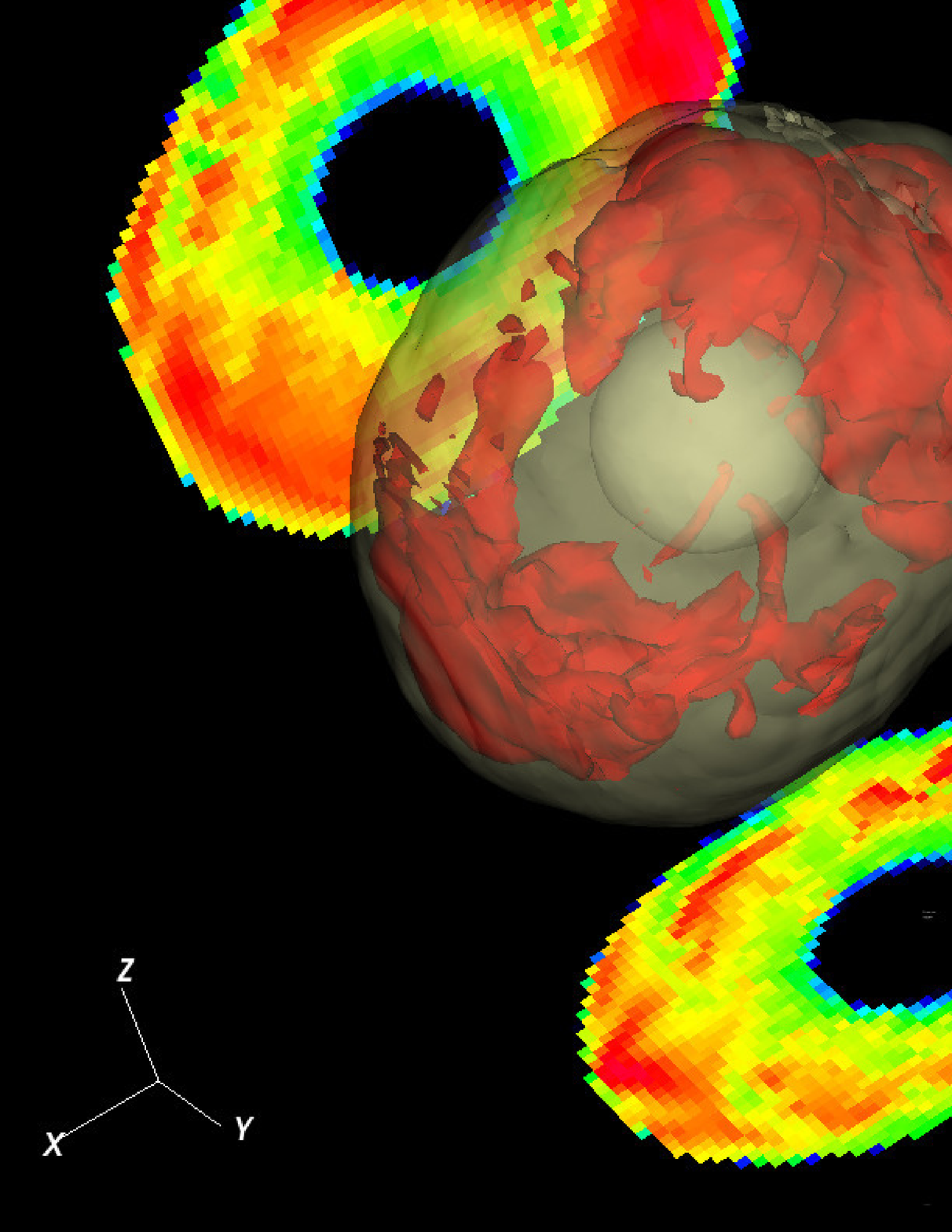}
          \includegraphics[clip,width=70mm]{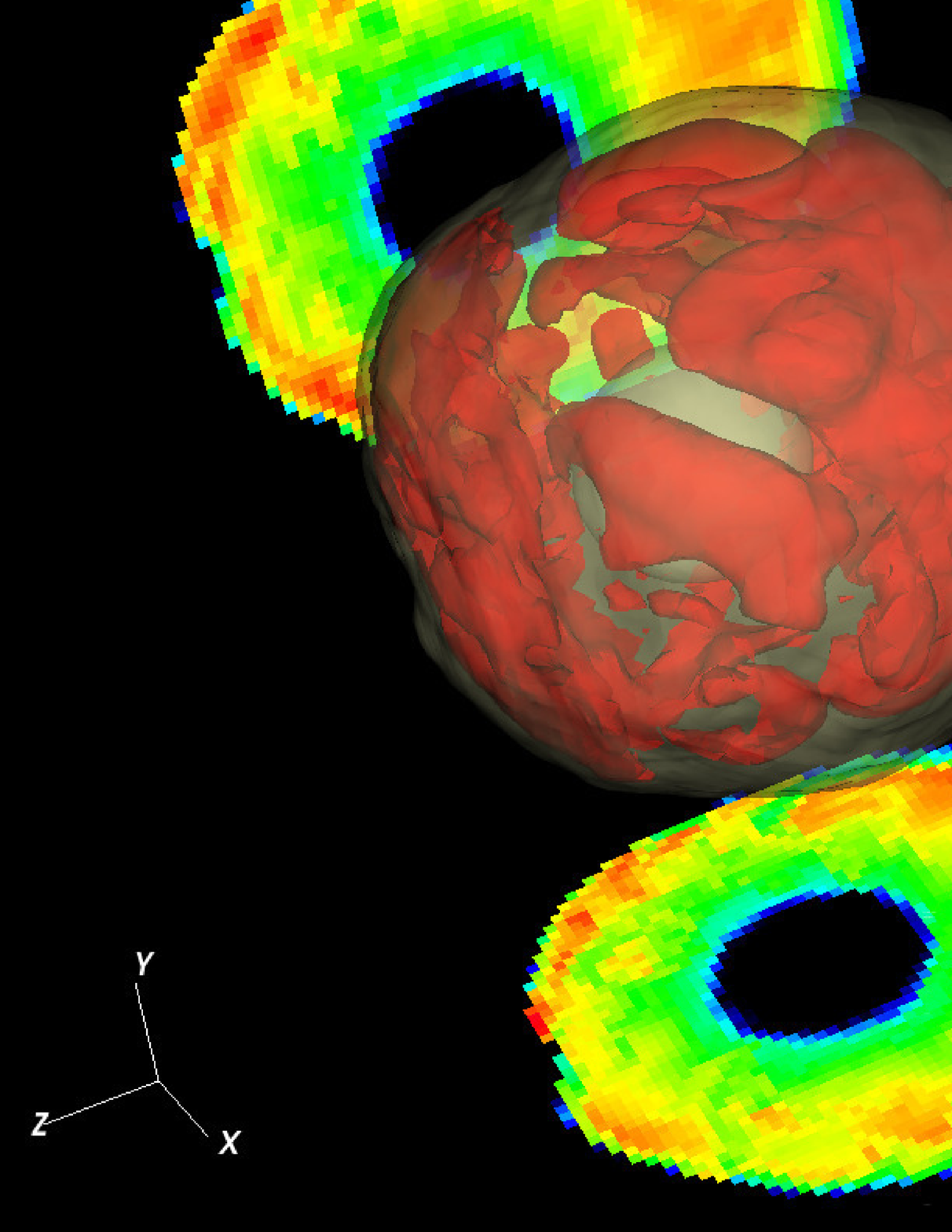}
          \includegraphics[clip,width=70mm]{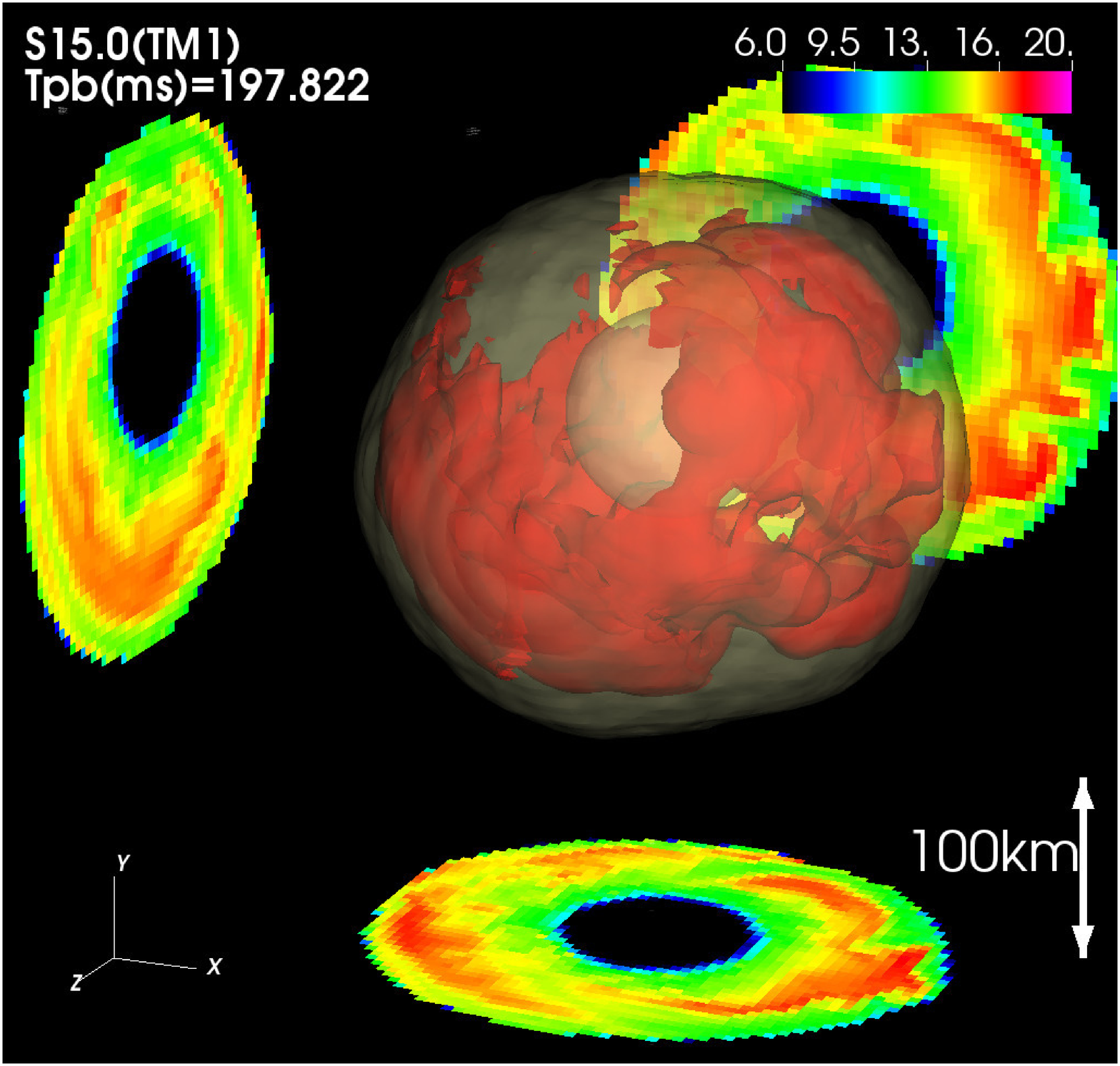}
        \end{center}
      \end{minipage}
      \begin{minipage}[t]{0.5\hsize}
        \begin{center}
          \includegraphics[clip,width=70mm]{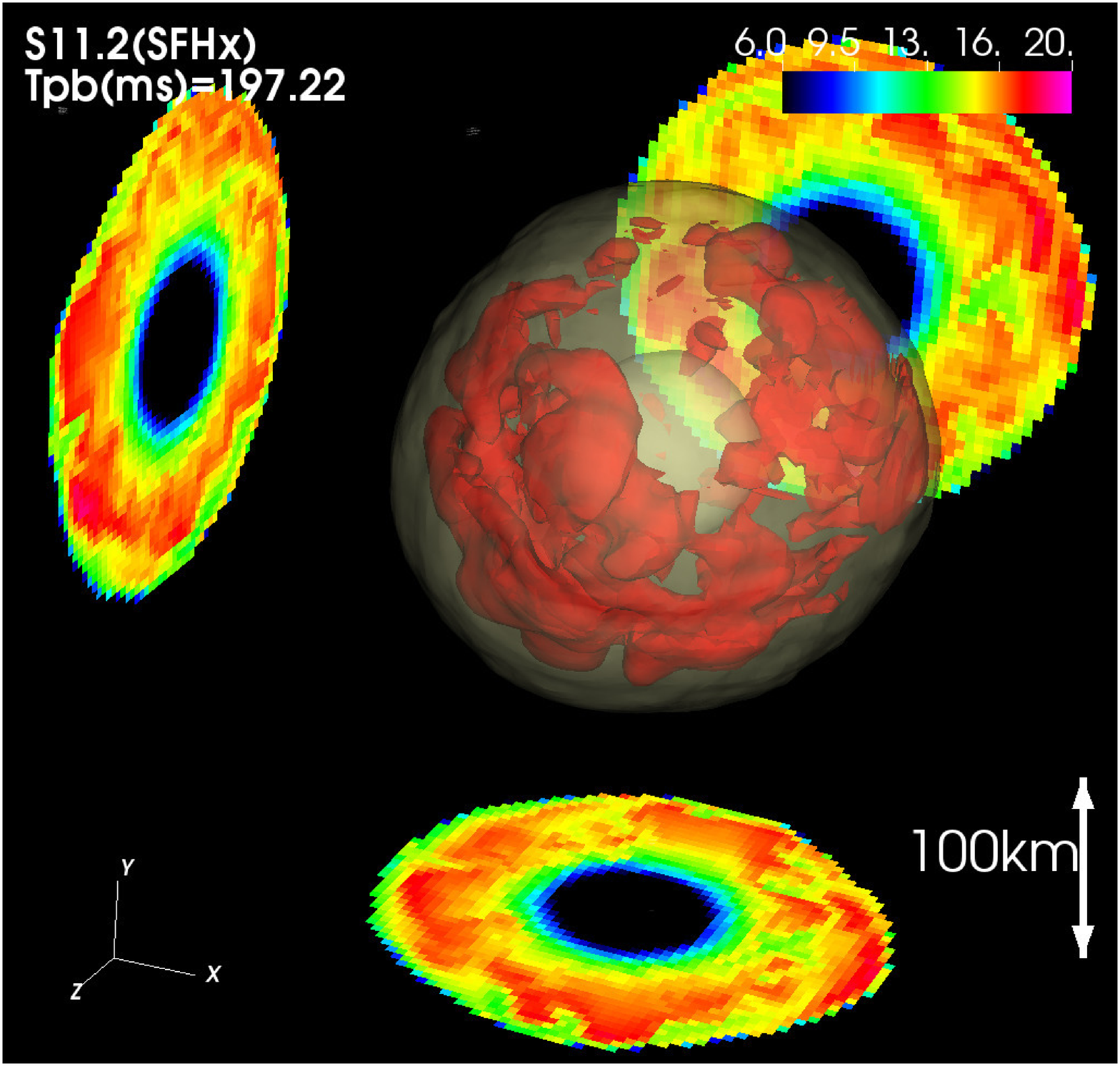}
          \includegraphics[clip,width=70mm]{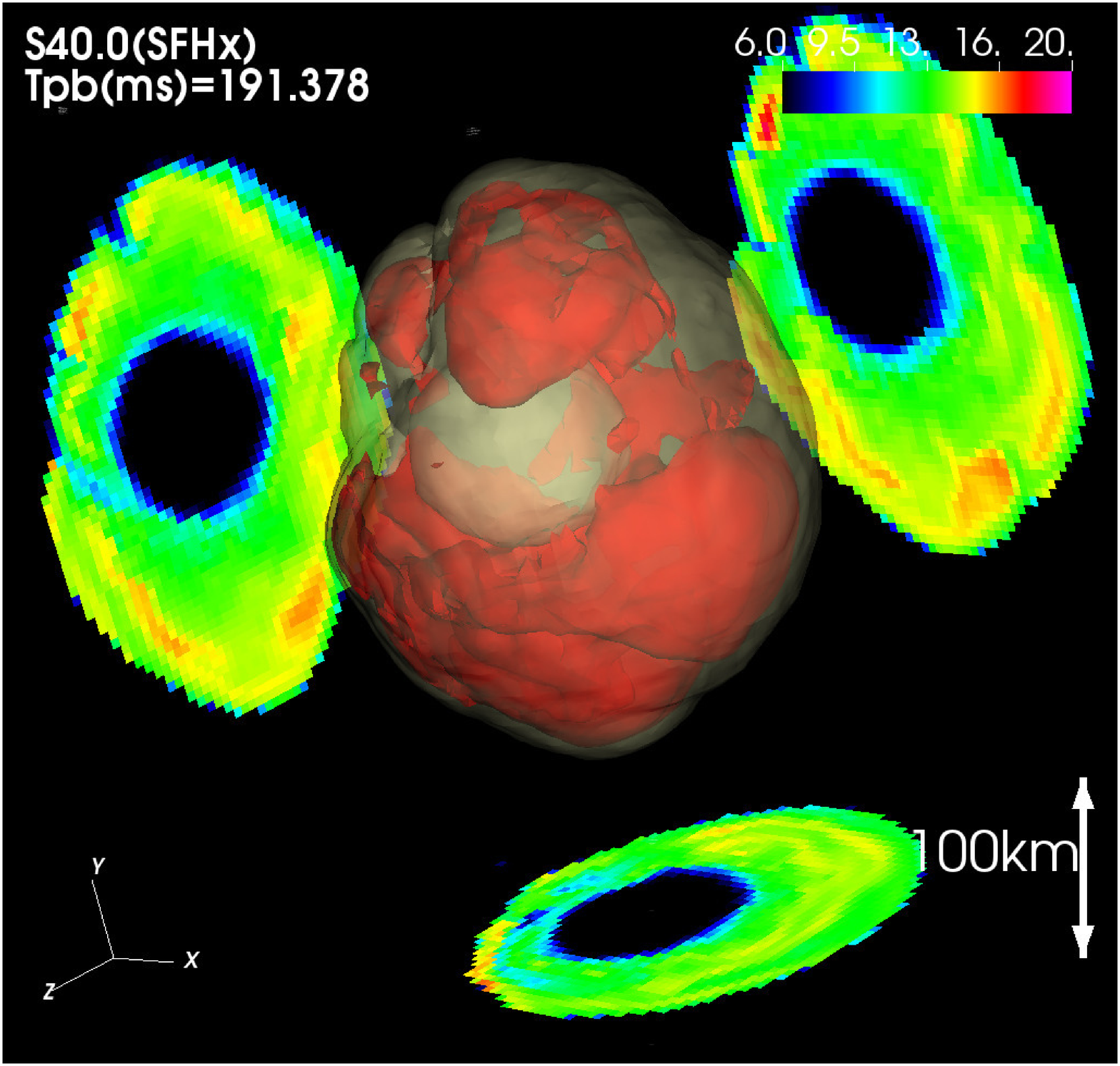}\\
        \end{center}
      \end{minipage}
    \end{tabular}
    \caption{Snapshots showing hydrodynamics features of all the computed models 
at representative time snapshots. Shown are the isentropic surfaces 
for $s=7$ $k_{\rm B}$ baryon$^{-1}$ (transparent shell)
  and for $s=17$ $k_{\rm B}$ baryon$^{-1}$ (red bubbles) 
(from top to bottom, left column; S15.0(SFHx), S15.0(DD2), and S15.0(TM1), right column; S11.2(SFHx) and S40.0(SFHx)).
$T_{\rm pb}$ denotes the postbounce time.
The contours on the cross sections in the $x$ = 0, $y$ = 0, and $z$ = 0 planes are projected on the sidewalls.
The left column focuses on the EOS dependence.
Top left and right column show the progenitor mass dependence. 
  \label{f2}}
  \end{center}
\end{figure*}

Figure \ref{f1} compares evolution of the average (thick lines) and maximum 
(dash-dotted lines) shock radii for models with different EOSs (panel (a)) and
 with different progenitor masses (panel (b)), respectively. 
Before $T_{\rm pb}\sim150$ ms (panel (a)), the average and 
maximum shock radii are smallest for SFHx (red thick line), followed 
 in order by DD2 (turquoise line) and TM1 (black line), which is 
 exactly the same as the stiffness of the EOS (SFHx:softest, TM1:stiffest).
 However, after the non-linear phase sets in ($T_{\rm pb} \ga 150$ ms) 
when neutrino-driven convection and the SASI develop vigorously with time,
the maximum shock radii of SFHx becomes biggest followed in order by DD2 and 
 TM1. This reversal of the maximum shock radius before and after the non-linear 
 phase is due to the more stronger growth of the SASI for softer EOS.
As previously identified \citep{Scheck08,Hanke13}, this is because the smaller 
shock radius and the more compact core ($M_{\rm cb}/R_{\rm cb}$ in Table 1) 
lead to more efficient advective-acoustic cycle, i.e., 
 the SASI activity \citep{Foglizzo02,Foglizzo06}. Figure \ref{f2} visually supports 
this, 
where the large-scale shock deformation is most clearly seen for S15.0(SFHx) 
(top left panel), whereas the shock deformation is more modest for S15.0(DD2)
 (middle left panel) and for S15.0(TM1) (bottom left panel).

Panel (b) of Figure \ref{f1} compares the shock radii for the different 
 progenitors with the same EOS (SFHx). S11.2(SFHx) shows the largest shock radii
 (average/maximum, blue lines) before $T_{\rm pb} \lesssim 160$ ms. This is because 
prompt convection develops 
 much more strongly for S11.2(SFHx). As consistent with \citet{BMuller13}, 
this is because the prompt shock 
propagates rapidly due to the smaller mass accretion rate.
 Prompt convection is observed by formation of small-scale convective
 motions behind the roundish stalled shock (see the top right panel 
 of Figure \ref{f2}). The absence of clear SASI activity of this model is in accord with 
 \citet{Hanke13,bernhard16} where the 11.2 $M_{\odot}$ star was used in 
their self-consistent 3D models (but with different EOSs used). 
In the panel (b), the average shock radius is slightly more compact
 for S15.0(SFHx) than S40.0(SFHx) in the linear phase ($T_{\rm pb} \lesssim 150$ ms). 
 In both S15.0(SFHx) and S40.0(SFHx), the SASI activity was similarly observed
 in the non-linear phase (bottom panel of Figure \ref{f2}), whereas 
 the maximum shock radius is generally bigger for S15.0(SFHx). We ascribe 
 this to the high SASI activity of S15.0(SFHx) compared to S40.0(SFHx) 
 predominantly due to the more compact core ($M_{\rm cb}/R_{\rm cb}$ in Table 1). 

\section{GW signatures}
\label{sec4}
\begin{figure*}[htbp]
  \begin{center}
    \begin{tabular}{cc}
      \begin{minipage}[t]{0.48\hsize}
        \begin{center}
          \includegraphics[clip,width=68mm,angle=-90.]{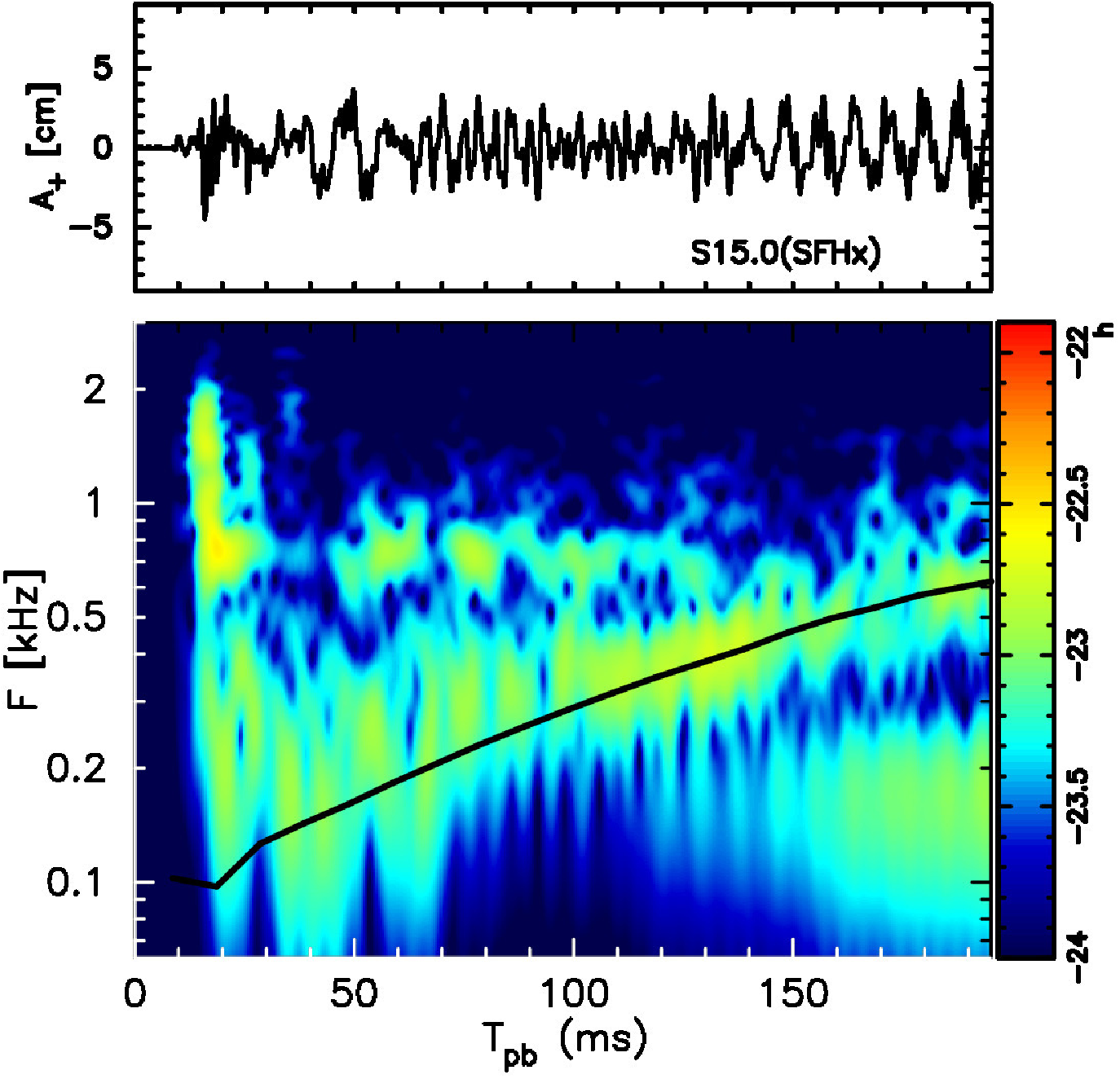}
          \includegraphics[clip,width=68mm,angle=-90.]{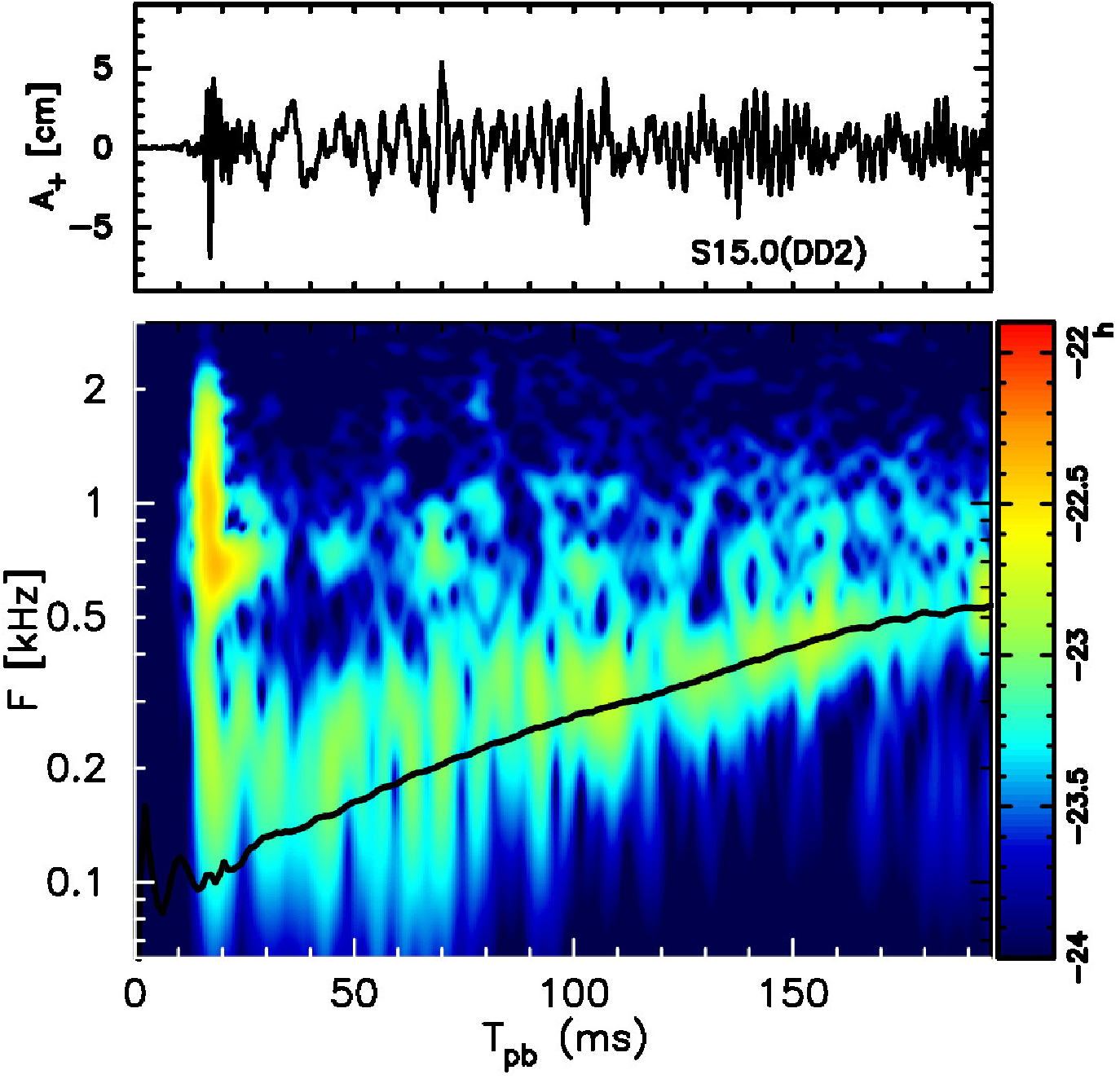}
          \includegraphics[clip,width=68mm,angle=-90.]{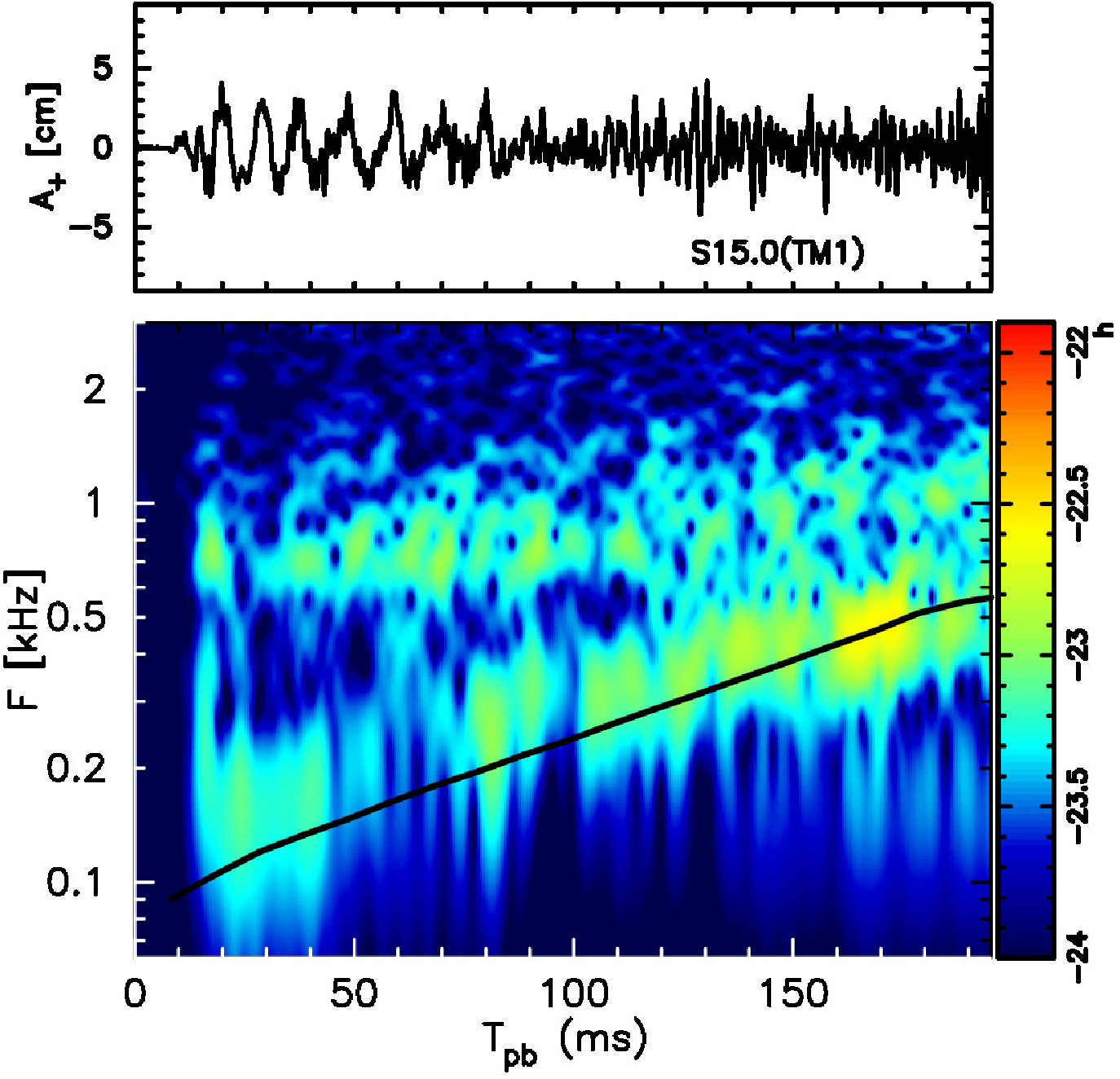}
        \end{center}
      \end{minipage}
      \begin{minipage}[t]{0.48\hsize}
        \begin{center}
          \includegraphics[clip,width=68mm,angle=-90.]{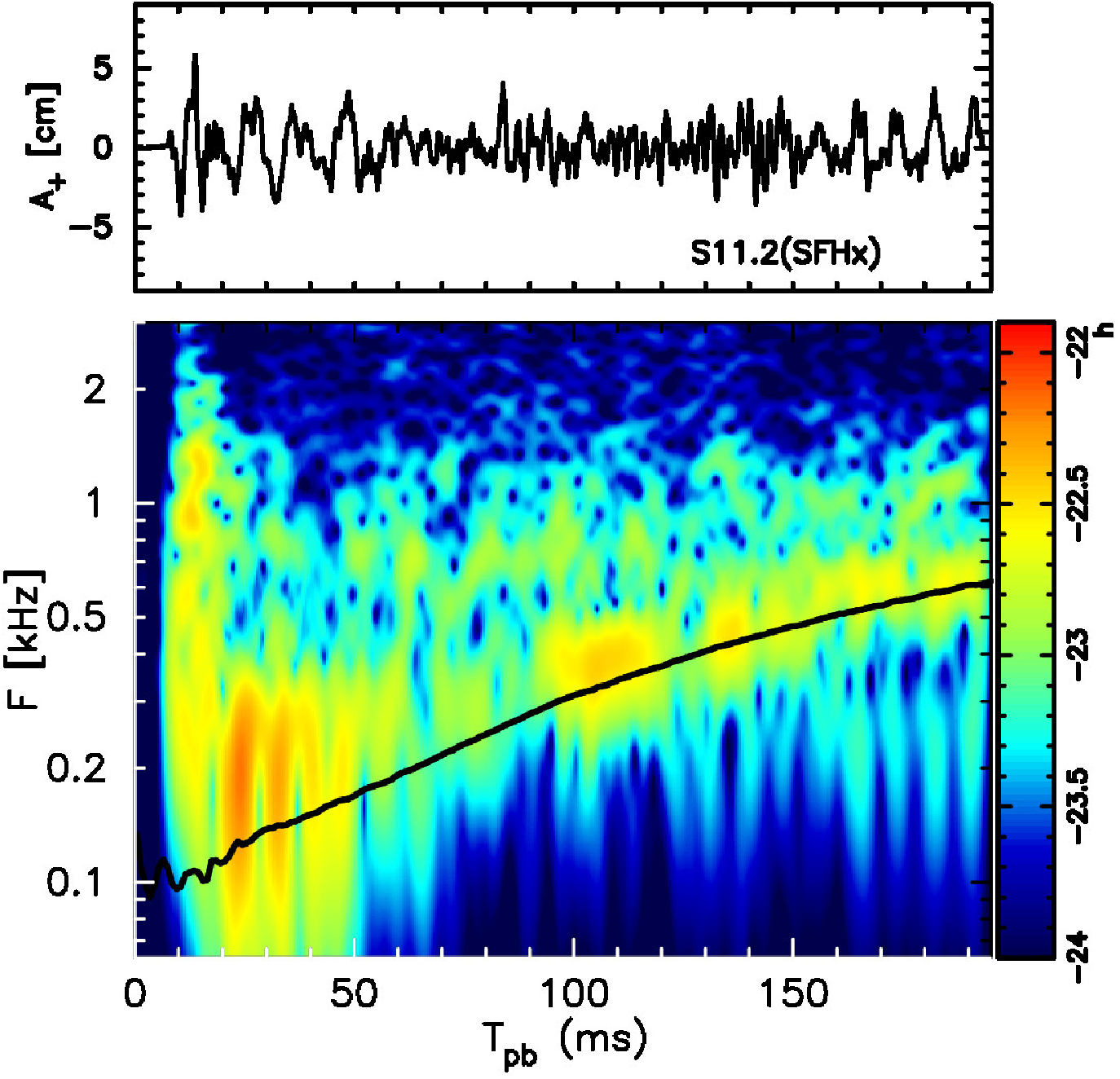}
          \includegraphics[clip,width=68mm,angle=-90.]{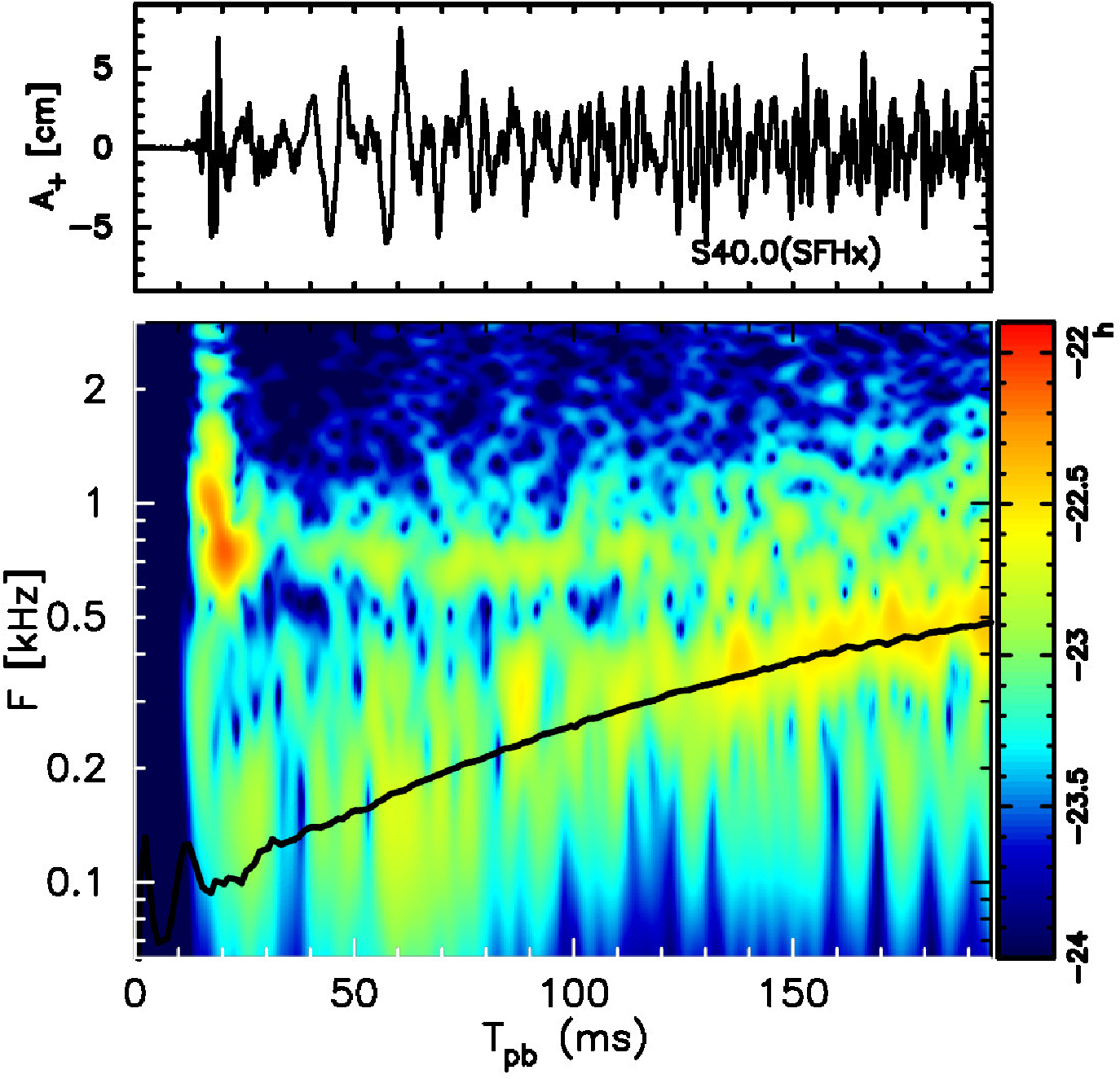}\\
        \end{center}
      \end{minipage}
    \end{tabular}
    \caption{In each set of panels, we plot ({\it top})
the GW amplitude of plus mode $A_+$ [cm] and ({\it bottom}) 
the characteristic wave strain in the frequency-time domain $\tilde h$ in a 
logarithmic scale that is overplotted by the analytical GW frequency $F_{\rm peak}$ (black line) of the PNS $g$-mode oscillation \citep{Marek09,BMuller13,CerdaDuran13}.
  We note that SFHx ({\it top left}) is the softest EOS followed in order by 
 DD2 ({\it middle left}), and TM1 ({\it bottom left}), respectively. The top and middle right and 
panels are for S11.2(SFHx) and S40.0(SFHx), respectively.
  \label{f3}}
  \end{center}
\end{figure*}


In this section, we summarize how the hydrodynamics features 
in Section \ref{sec3} impact the GW emission.

  Figure \ref{f3} shows time evolution of the GW amplitude (only plus mode $A_+$
  and extracted along positive $z$-axis, black line) in the top
 panels and the characteristic wave strain in the frequency-time domain
($\tilde h(F)$, e.g., Eq. (44) of \citet{KurodaT14}) in the bottom
 ones. Here $F$ denotes the GW frequency.  The top panels show a consistent GW 
 behavior as previously identified in self-consistent models of 
\citet{BMuller13,KurodaT16ApJL,Andresen17,Yakunin15}. 
After bounce, the wave amplitude deviates from zero with low/high-frequency and 
relatively large spikes until $T_{\rm pb} \sim 50$ ms. This is due to prompt convection. 
The GW from the prompt convection is shown to be biggest for S11.2(SFHx) 
(middle right panel). This comes from the vigorous prompt 
convection activity of this model as already mentioned in Section \ref{sec3}
(e.g., Figure \ref{f1}(b)). It is consistent with \cite{BMuller13}, who 
 also showed a factor of $\sim1.5$ 
larger GW amplitude from prompt convection in the 11.2 $M_{\odot}$ model (G11.2)
compared to that of $15 M_{\odot}$ model (G15).

After prompt convection, no common features in the waveforms 
can be found among the models reflecting stochastic nature of the postbounce GWs.
However, guided by the black line (Figure \ref{f3}), one can see a relatively 
narrow-banded spectrum for all the models that shows an increasing trend in its peak 
frequency. In addition to this PNS $g$-mode contribution \citep{BMuller13,Murphy09,CerdaDuran13}, the SASI-induced low-frequency component is clearly seen 
for S15.0(SFHx) (e.g., the excess around $100 \lesssim F \lesssim150$ Hz 
 at $T_{\rm pb} \gtrsim 150$ ms in the spectrogram (top left panel)). Note that 
 this is also observed in \citet{Andresen17}. 

\begin{figure}[htpb]
\begin{center}
  \includegraphics[width=65mm,angle=-90.]{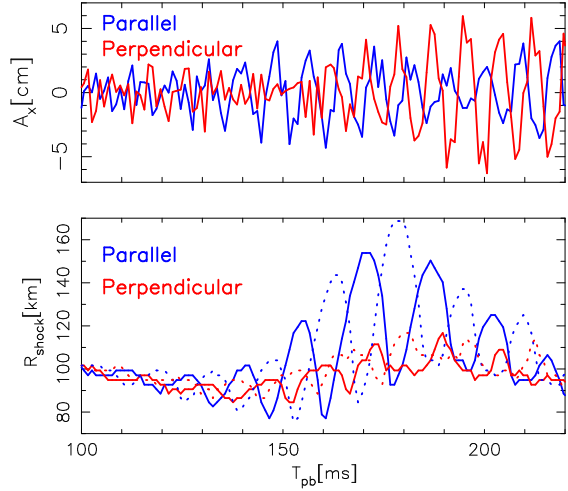}
  \caption{Time evolutions of the GW amplitudes (cross mode, top) and shock positions (bottom).
  The color represents that the observer direction is parallel to the sloshing SASI
 axis (blue line) or perpendicular to the sloshing axis (red line, for a given 
 azimuthal direction) in the top panel.
In the bottom panel, the shock positions are measured along the two lines of sight
 with the same color in the top panel, where the solid and dotted curves
 correspond to the shock position at the nearest or farest to the observer
 along the line of sight, respectively.
  \label{f10}}
\end{center}
\end{figure}
So far, we show results only for one representative observer direction (along positive $z$-axis) which is not a special direction relative to the SASI motion.
\cite{Tamborra14} showed that the time modulation in neutrino signal has a dependence on the observer angle relative to
the (sloshing) SASI motion.
According to their results, neutrino detection rate is significantly larger and also the time modulation is more clearly seen when the observer is along the axis of sloshing motion.

As we have discussed in the previous Sec. \ref{Sec:Overview of Hydrodynamics Features} \citep[see also][]{KurodaT16ApJL},
  some of our models show vigorous sloshing SASI motion.
  To see the observer angle dependence on the GW, we plot the GW amplitudes (only for the cross mode, top)
  and shock positions (bottom) as a function of post bounce time in Fig. \ref{f10} for model S15.0(SFHx).
  To plot this figure, we first determine two lines of sight.
  One is parallel to the sloshing axis and the other is an arbitrary direction but perpendicular to the sloshing axis. 
Then the color in Fig. \ref{f10} represents that the observer direction is parallel (blue)
or perpendicular (red) to the sloshing axis in the top panel.
In the bottom panel, the shock positions are measured along these two lines of sight with the same color notation in the top panel.
Two shock positions along each line of sight, near and far side to the observer, are plotted by different line styles, solid (near) and dotted (far), respectively.
As a reference, the observer directions are $(\theta,\phi)\sim(135^\circ,0^\circ)$ and $\sim(45^\circ,0^\circ)$
for the parallel and perpendicular directions, respectively, at $T_{\rm pb}\sim180$ ms for S15.0(SFHx).

As one can see from the bottom panel, the shock position oscillates largely along the sloshing axis (blue lines)
with nearly the opposite phase between the solid and dotted lines.
On the other hand the red lines show significantly smaller deviations.
After the sloshing motion reaches its maxima at $T_{\rm pb}\sim180$ ms, the GW emitted toward the perpendicular direction
reaches $\sim5$-6 cm at $180\la T_{\rm pb}\la 200$ ms.
In the meantime, the GW amplitude reaches merely $\sim2$ cm along the parallel direction.
Thus, contrary to the neutrino emission, the GW emission is stronger toward the orthogonal direction to the sloshing motion.
 This is analogous to the stronger GW emission toward the equatorial plane in the rotating progenitor model at bounce.

\begin{figure}[htpb]
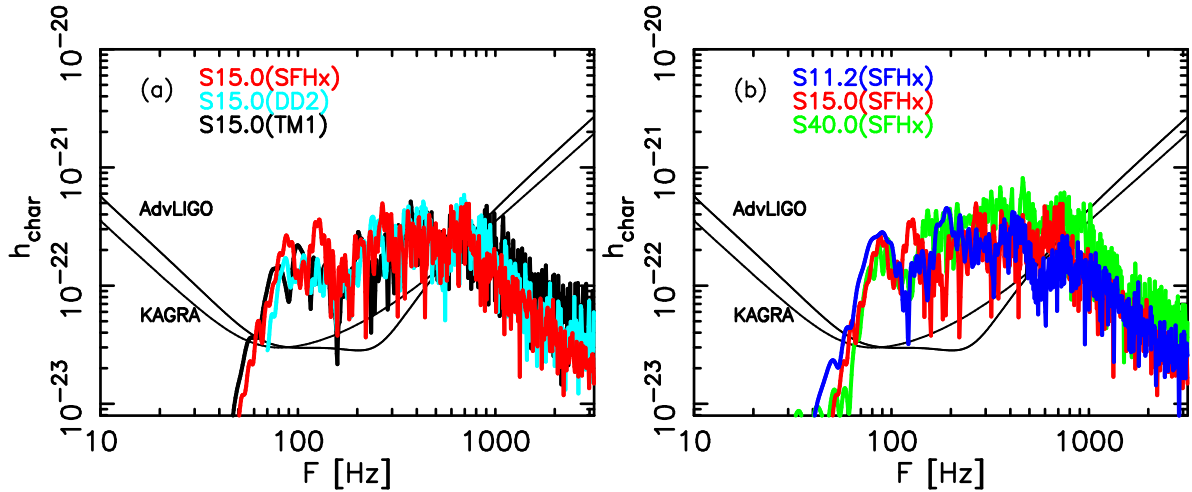

\begin{center}
  \includegraphics[width=65mm,angle=-90.]{f13.eps}
    \includegraphics[width=65mm,angle=-90.]{f14.eps}
  \caption{Same as Figure \ref{f1}, but for the characteristic GW strain spectra
 for a source of distance at 10 kpc. We estimate the spectra for 
the time integration of the GW energy in the range of $0\le T_{\rm pb}\le200$ ms.
Solid thin black curves denote the sensitivity curves of LIGO \citep{Harry10} and KAGRA \citep{Aso13}.
  \label{f4}}
\end{center}
\end{figure}

Regarding the EOS dependence, the GW spectrum extends to higher frequency 
for our model with the stiffest EOS (S15.0(TM1), black line in the left panel of Figure 
\ref{f4}), whereas the GW spectrum for the softest EOS (S15.0(SFHx), red line) 
 concentrates more in the lower-frequency domain. Note that an excess 
 around $100 \lesssim F \lesssim 200$ Hz in the spectrum 
 of S15.0(SFHx) corresponds to the SASI-induced GW emission mentioned above.

As for the progenitor dependence, S40.0(SFHx) shows stronger GW emission 
 compared to S11.2/15.0(SFHx) (e.g., bottom panel of Figure \ref{f3}).
 In fact, the right panel of Figure \ref{f4} (green line) shows that 
the GW spectrum dominates over that of the other models 
 over the wide frequency range.
For this model, the signal to noise ratio reaches $\sim10$ around the best 
sensitivity around $F\sim100$ Hz 
and a Galactic event could be likely to be detectable. But, in order to 
 discuss the detectability of the signals more quantitatively, one needs a
 dedicated analysis (e.g, \citet{Hayama15,Powell16,Gossan16}), which is beyond the scope of this work.



\section{Correlation between GW and Neutrino Emission}
\label{sec5}
 In this section, we present a correlation analysis between the GW (Section \ref{sec4}) and the neutrino signals.

\begin{figure*}[htbp]
  \begin{center}
    \begin{tabular}{cc}
      \begin{minipage}[t]{0.48\hsize}
        \begin{center}
          \includegraphics[clip,width=68mm,angle=-90.]{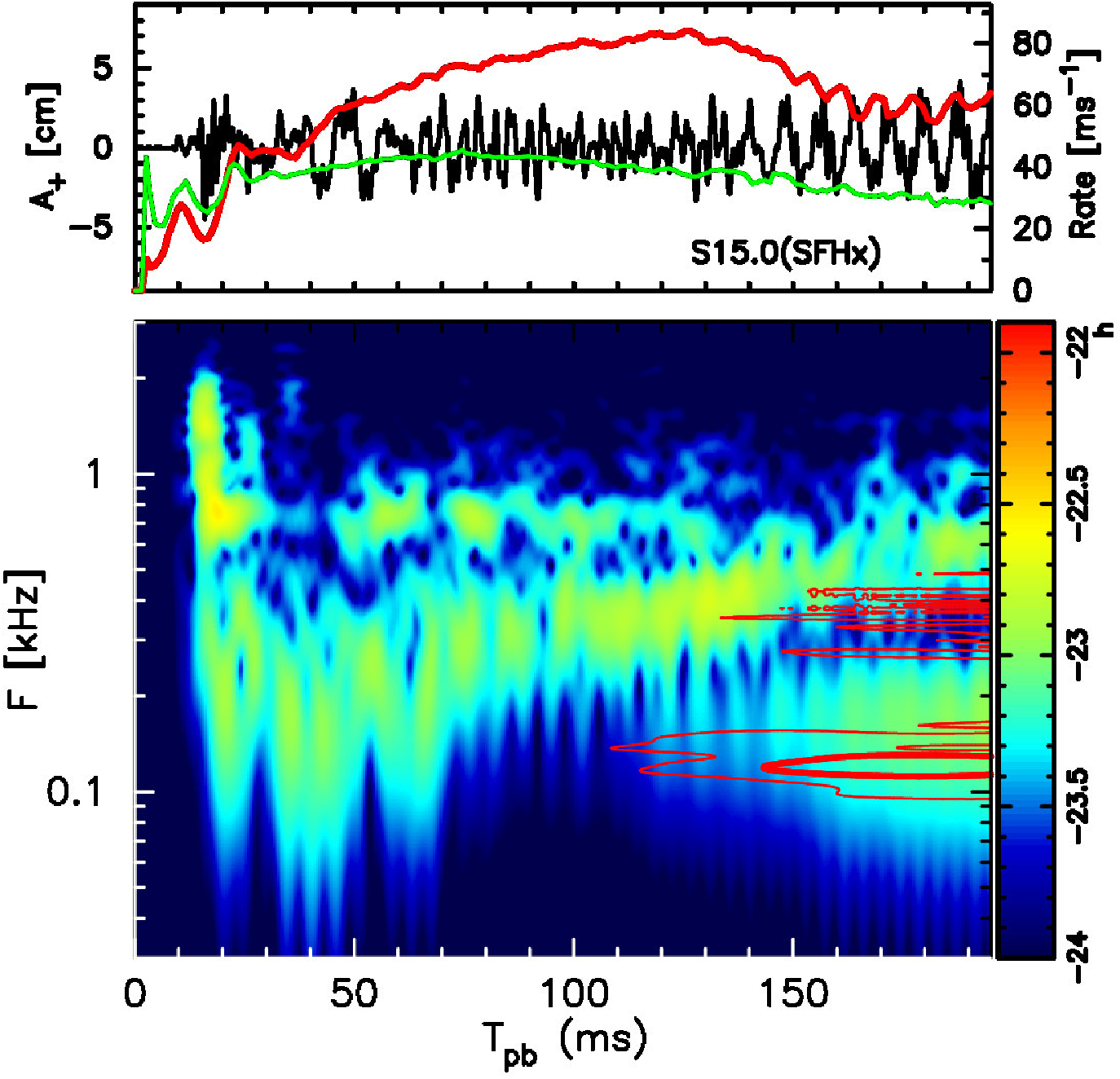}
          \includegraphics[clip,width=68mm,angle=-90.]{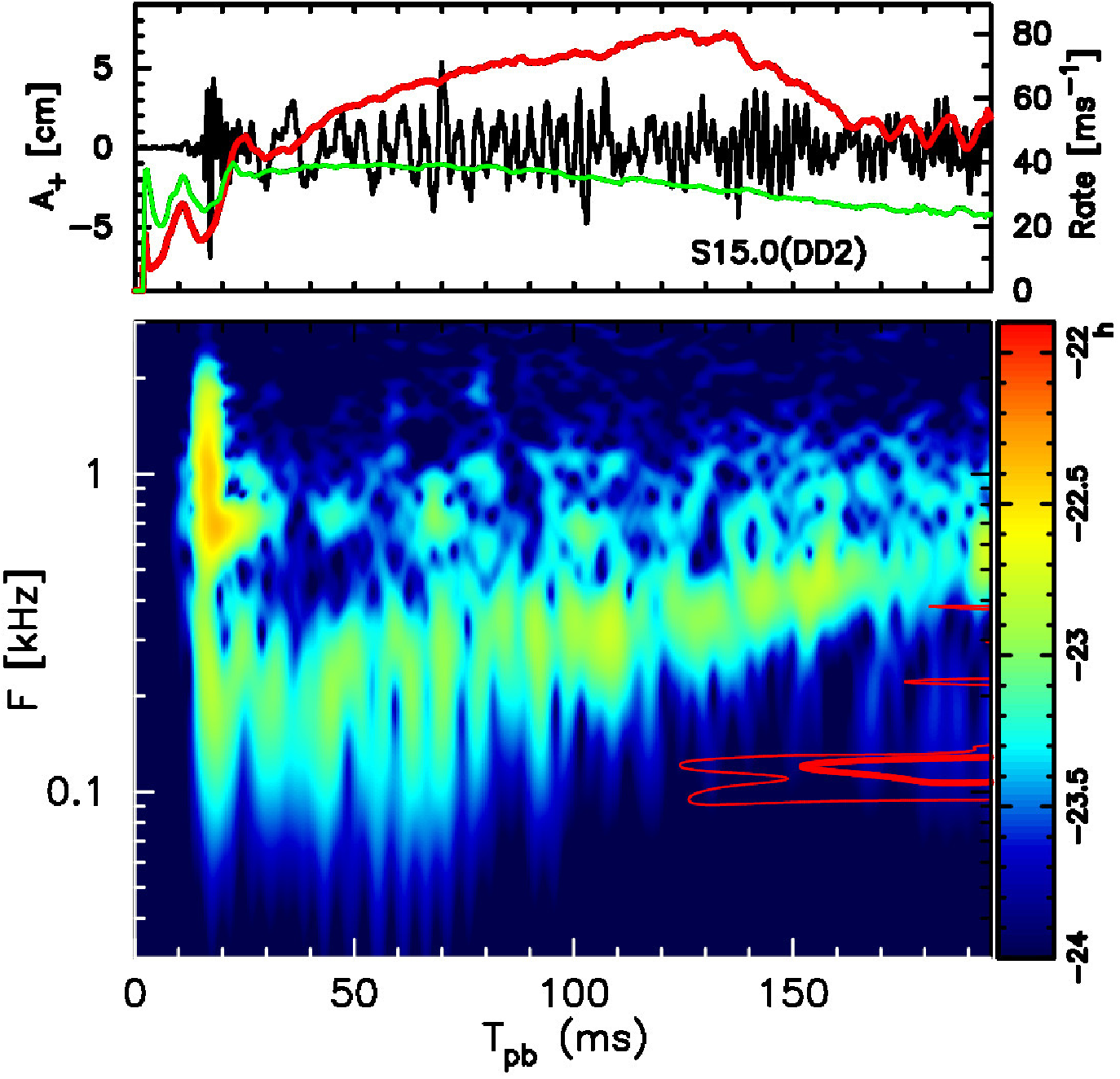}
          \includegraphics[clip,width=68mm,angle=-90.]{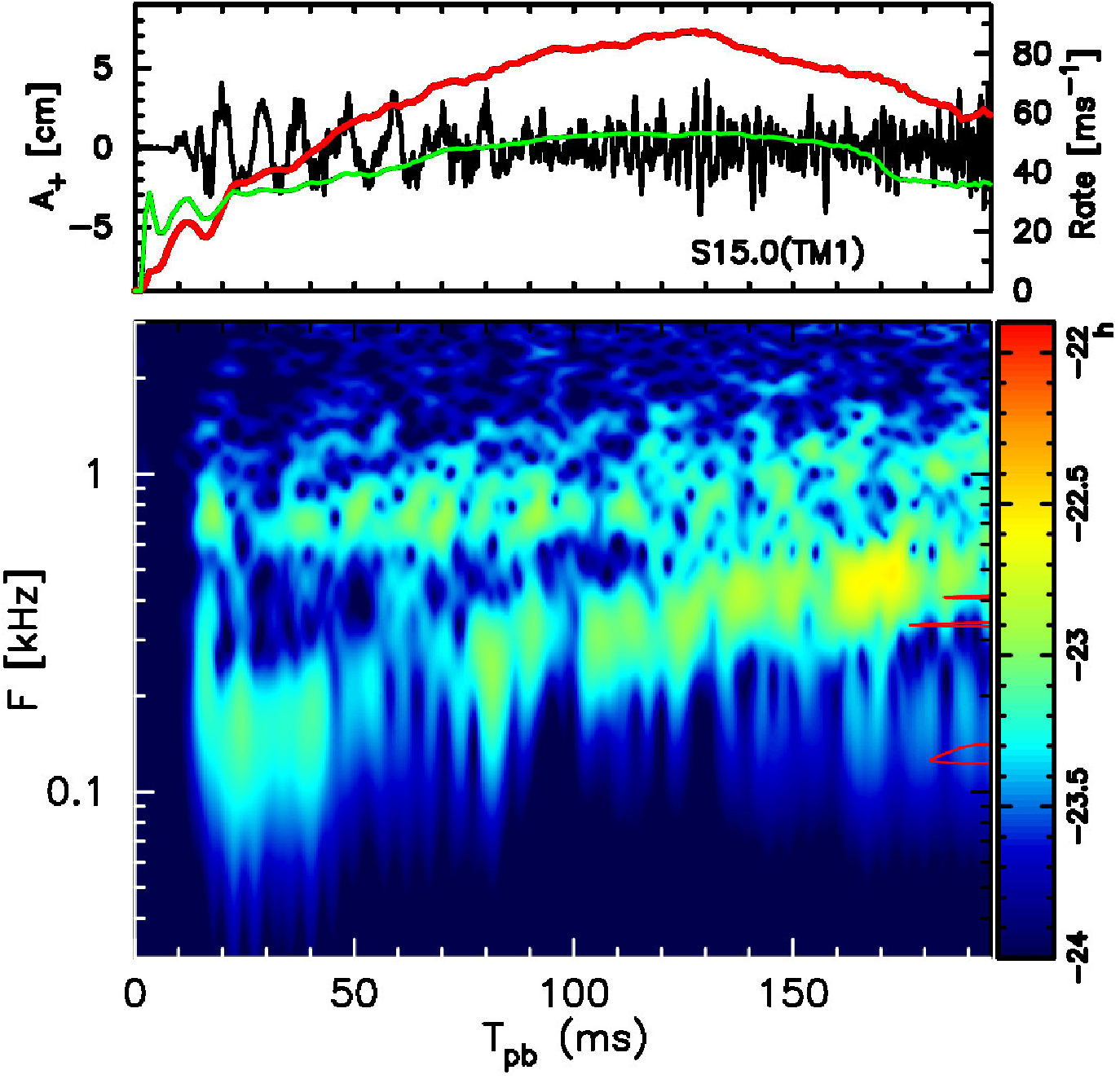}
        \end{center}
      \end{minipage}
      \begin{minipage}[t]{0.48\hsize}
        \begin{center}
          \includegraphics[clip,width=68mm,angle=-90.]{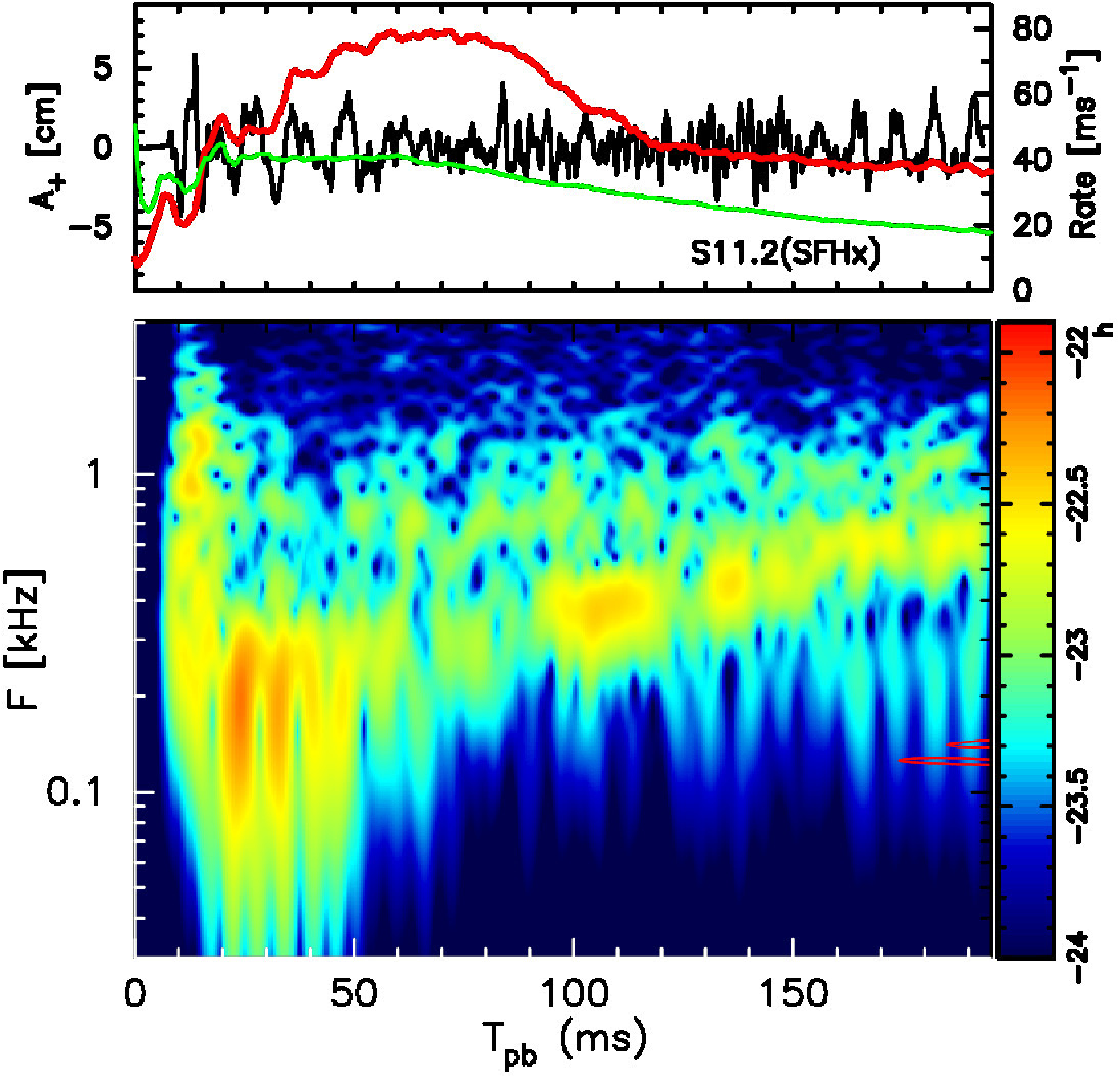}
          \includegraphics[clip,width=68mm,angle=-90.]{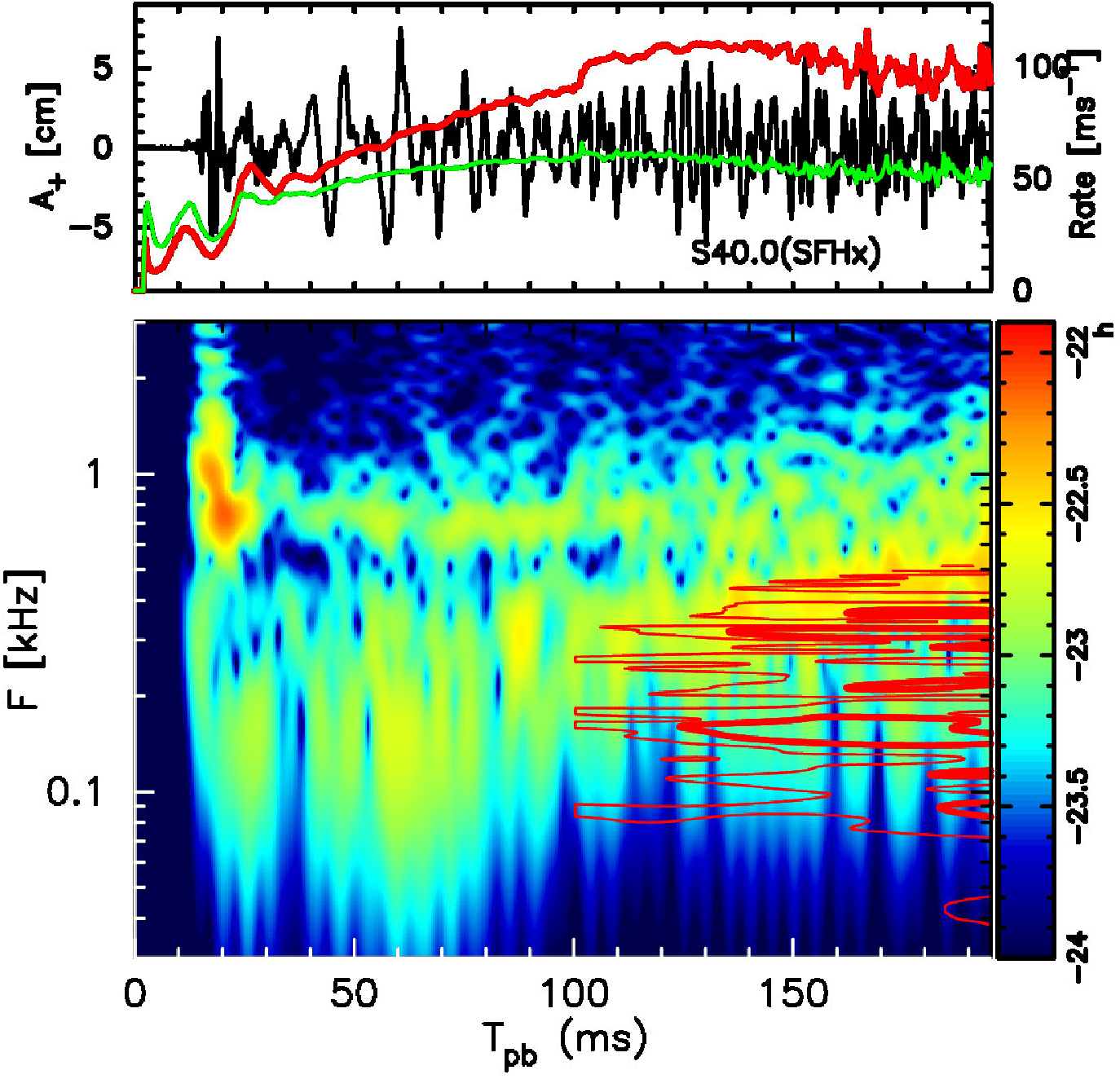}\\
        \end{center}
      \end{minipage}
    \end{tabular}
    \caption{For each model, the top panel shows the neutrino event rate 
$N_\nu\,[\rm ms^{-1}]$ (red and green lines are for $\bar\nu_e$ and $\nu_x$, respectively) for Hyper-K and the 
GW amplitude $A_+$ [cm] (black line), whereas in the bottom panel we plot contours 
(red curves, only for $T_{\rm pb}\ge100$ ms) of the anti-electron type neutrino spectra that are 
superimposed on the color-coded GW spectrum. 
The observer's direction is fixed along the $z$-axis for a source at 
a distance of $D=10$ kpc.
\label{f5}}
  \end{center}
\end{figure*}

For all the computed (five) models, we plot in Figure \ref{f5} the expected 
neutrino event rate ($N_\nu$ [ms$^{-1}$], red line) for Hyper-K (fiducial mass 440 kton, 
\citet{Hyper-K2016})
and the GW amplitude $A_+$ (black line) in the top panel. In the bottom panel,
 the contours (red curves) correspond to
 the Fourier-decomposed anti-electron type neutrino event rates (two arbitrary chosen values of $dN_\nu/dF=0.4$ (thin red line) and 
0.8 (thick red line), only for $T_{\rm pb}\ge100$ ms) that is superimposed on the GW spectrograms.
As similar to Fig. \ref{f3}, the observer direction 
for both neutrinos and GWs is fixed along the $z$-axis with 
a source distance of $D=10$ kpc.
Following the methods in Appendices A and B of \citet{Tamborra14ApJ},
we estimate the expected neutrino event rate from our 3D models,
 where the flux-projection effects are also taken into account.
As consideration of collective neutrino oscillation is apparently 
 beyond the scope of this work (e.g., \citet{duan10,mirizzi16,chakra16} for reviews),
we show two extreme cases where the detector measures the original $\bar\nu_e$ (red line) or $\nu_x$ (green line) flux.
 The latter case corresponds to the complete flavor conversion through the Mikheyev-Smirnov-Wolfenstein (MSW) effect \citep{MSW78,MSW85}.

Among the models in Figure \ref{f5}, the top left panel (S15.0(SFHx)) shows 
 clearest overlap between the neutrino modulation (see red contours in the 
 spectrogram) and 
the GW modulation at $T_{\rm pb} \gtrsim 150$ ms in the frequency range of 
 $F \sim 100$-150 Hz. For S15.0(TM1) with stiffest EOS (middle left panel), 
  the overlap between the quasi-periodic neutrino and GW signals
 can be marginally seen only at higher frequency range $F\sim400$-500 Hz after 
$T_{\rm pb}\sim150$ ms, which is significantly weaker compared to S15.0(SFHx).
Comparing S15.0(TM1) with S15.0(DD2) (top right panel), one can see a
 clearer quasi-periodic oscillation in the neutrino event rate for softer EOS
(top right panel), although there is little correlation between the GW and neutrino signal in the spectrogram. 
In the smallest mass progenitor of S11.2(SFHx), we do not find any 
remarkable simultaneous oscillation of the neutrino and GW signals.
For this model, the neutrino event rate becomes smallest among the five models and 
shows little time modulation (after $T_{\rm pb}\sim100$ ms), which is consistent with
 \citet{Tamborra14}.
On the other hand, the most massive progenitor of S40.0(SFHx) has a 
 largest overlap in the spectrogram (red contours) over the wide
 frequency range $50\lesssim F \lesssim 500$ Hz.

 When the complete flavor conversion between $\bar{\nu}_e$ and ${\nu}_x$ is assumed
(green line at $T_{\rm pb}\ga150$ ms of model S15.0(SFHx) in Fig. \ref{f5}),
 the time modulation is significantly suppressed as already reported in 
\cite{Tamborra13}.
 This is because that the neutrino spheres of heavy-lepton neutrinos 
 are located much deeper inside compared to those of anti-electron neutrinos.
Consequently they are less affected by the SASI activity and the 
correlation between the GW and the neutrino event rate becomes weaker 
in the case of the complete flavor swap.

\begin{figure*}[htpb]
\begin{center}
 \includegraphics[width=50mm,angle=-90.]{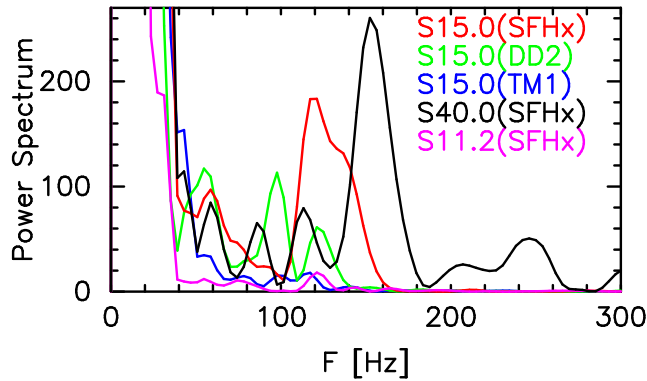}\\
  \caption{The power spectrum of the IceCube event rate for the time interval of $100\le T_{\rm pb}\le200$ ms.
 \label{f6}}
\end{center}
\end{figure*}

We plot in Figure \ref{f6} power spectra of the neutrino events in IceCube \citep{IceCube} to see impacts of the EOS and the progenitor. 
 A pronounced peak is seen around $\sim120$ Hz in S15.0(SFHx) (red line),
 which is absent for other S15.0 models with weak SASI activity (green and blue lines).
 This is again consistent with \cite{Tamborra13,Tamborra14}.
 The absence of the SASI signature of the 11.2 $M_{\odot}$ model is in line with 
\cite{Tamborra13}.
S40.0(SFHx) that has a relatively high compactness parameter 
(Table \ref{tb:BounceProfile}) exhibits a SASI 
activity and shows a peak at $F\sim160$ Hz.
 In addition to the biggest peak, some secondary peaks are also seen on the black line
 as well as in other models, e.g., at $F\sim60$ Hz on the red line.
In \cite{Tamborra13}, these secondary peaks are hard to see in most of the employed 
progenitors except for the $20 M_{\odot}$ model.
We consider that this difference might be partially due to our simplified
transport scheme, where the neutrino matter coupling is controlled via several 
parameters (see \citet{KurodaT12} for more details).
Because of this, our neutrino signals may change more sensitively in response
 to the matter motion compared to those obtained in CCSN models with 
 more sophisticated neutrino transport. For example, during the prompt convection 
phase ($T_{\rm pb}\lesssim50$ms), our neutrino event rate shows an oscillatory behavior
 (see red/green line in every top panel in Fig. \ref{f5}) which is not seen 
in \cite{Tamborra13}. To clarify this, we need to perform 3D-GR simulations 
 with more elaborate neutrino transport scheme
 which is, unfortunately, computationally unaffordable at this stage.

\begin{figure}[htpb]
\begin{center}
\includegraphics[width=70mm,angle=0.]{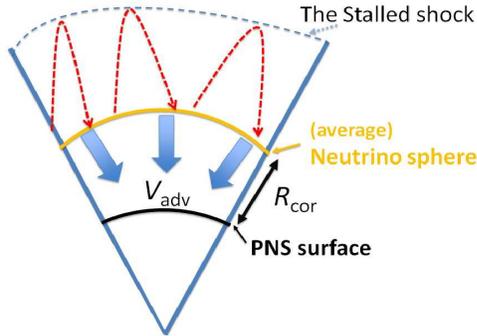}
\caption{Schematic drawing to illustrate the different radial positions 
of SASI-induced neutrino and GW emission in the postbounce core. Below the 
stalled shock (dashed blue line, labeled as ``The stalled shock''), non-spherical 
 flows (dashed red line with arrow) hit first the (average) neutrino sphere then 
 penetrates into the PNS core surface. $R_{\rm cor}$ represents the distance 
between the neutrino sphere ($\bar{\nu}_e$ in this case) and the PNS. $V_{\rm adv}$
 is the typical velocity of the downflows there.
\label{f7}}
\end{center}
\end{figure}

From Figures \ref{f5} and \ref{f6}, it has been shown that both of the  
SASI modulation frequency of the GW and neutrino signals is relatively close 
(i.e., in the range of 100 $\sim$ 200 Hz). Figure \ref{f7} illustrates how the two 
signals could be spatially correlated. In the figure, the SASI flows 
(red dashed arrows) advecting from the shock first excite oscillation 
in the neutrino signal at the (average) neutrino sphere. Afterward, it reaches 
to the PNS core surface (the blue thick arrows), leading to the modulation 
in the GW signal (see also \citet{KurodaT16ApJL} for the detailed analysis).
We can roughly estimate the time delay $\Delta T$ as follows. 
The radius of anti-electron type neutrino sphere is $R_{\bar\nu_e}\sim37$ km
 and the PNS core surface is $R_{\rm PNS}\sim15$ km (at $T_{\rm pb}=200$ ms for S15.0(SFHx)),
 then the correlation distance is $R_{\rm cor}=R_{\bar\nu_e}-R_{\rm PNS}\sim20$ km. 
 An angle-average accretion velocity at $R=40(20)$ km is
$V_{\rm adv}\sim -1\times10^8(-1\times10^7)$ cm s$^{-1}$ at $T_{\rm pb}=200$ ms,
 leading to $\Delta T$ of a few $10$ ms.

\begin{figure*}[htpb]
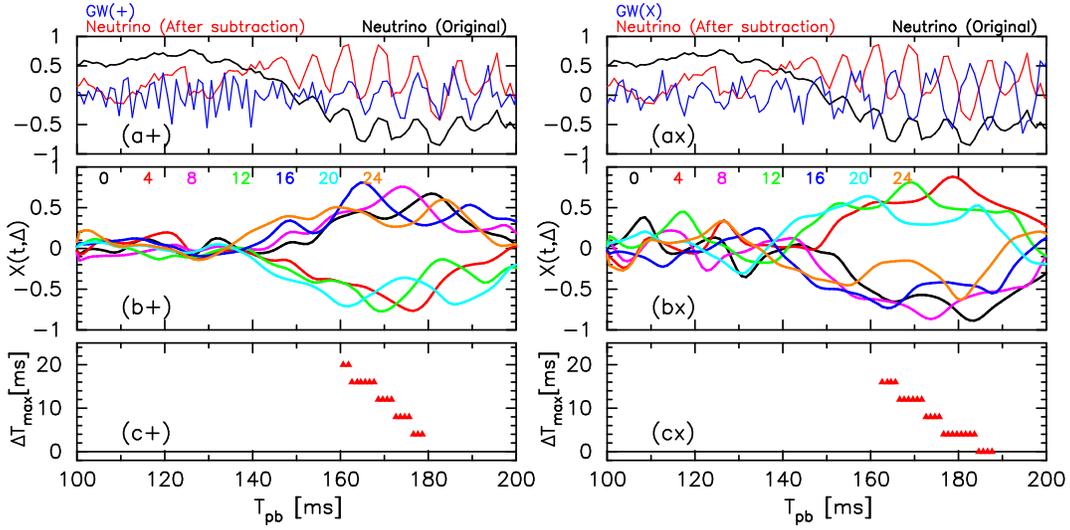

\begin{center}
  \includegraphics[width=70mm,angle=-90.]{f22.eps}
  \includegraphics[width=70mm,angle=-90.]{f23.eps}
  \caption{Top panels show the GW amplitude (blue line) 
either $+$ (left panel) or $\times$ polarization (right panel)
 and the neutrino event rate (black and red lines) in arbitrary units for S15.0(SFHx).
 For the red line, the monotonically time-changing component of the black line is subtracted
 ($T_{\rm pb}\la170$ ms) in order to focus on the SASI-induced modulation.
Same as the top panels, middle panels (b+/$\times$) show the correlation function 
$X(t,\Delta T)$ between the GW 
amplitude (blue line ({\it top})) and the event rate (red line ({\it top})) 
with several time delay $\Delta T$ (see text for definition) which is indicated
 in the upper left part as 0, 4, 8, 12, 16, 20, 24 [ms]. 
Bottom panels (c+/$\times$) show $\Delta T_{\rm max}$ that gives the delay-time 
with the maximum correlation in the middle panels.
 Note when we obtain $\Delta T_{\rm max}$, we set an arbitral threshold as $|X(t,\Delta T)|\ge0.7$ 
not to extract insignificant values.
  \label{f8}}
\end{center}
\end{figure*}

\begin{figure}[htpb]
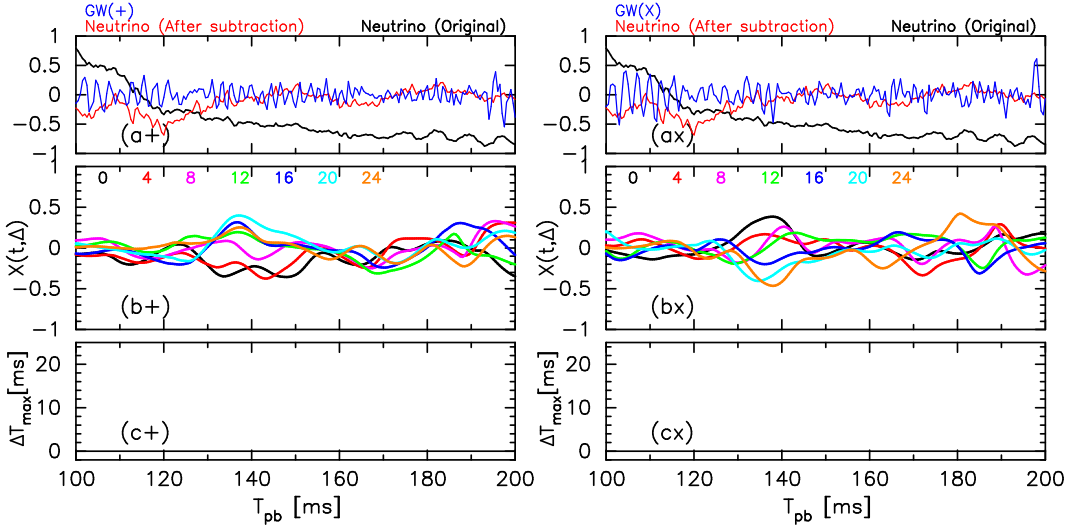

\begin{center}
  \includegraphics[width=70mm,angle=-90.]{f24.eps}
  \includegraphics[width=70mm,angle=-90.]{f25.eps}
  \caption{Same as Figure \ref{f8} but for S11.2(SFHx).
  \label{f9}}
\end{center}
\end{figure}

In order to estimate the correlation between the neutrino and GW signal more 
quantitatively, we evaluate the correlation function $X(t,\Delta T)$ in Figures 
 \ref{f8} and \ref{f9}. Note Figures \ref{f8} and \ref{f9} are for S15.0(SFHx) and 
S11.2(SFHx) showing highest and invisible SASI activity in this work, respectively.

The top panel of Figure \ref{f8} shows the GW amplitude (blue line) 
and the neutrino event rate (black and red lines) in arbitrary units.
In order to focus on the SASI-induced modulation, the red curve 
 is the event rate after the monotonicaly time-changing component
 is subtracted from the original curve (black line)\footnote{
     As one can see from the red line in each top panel in Fig. \ref{f5},
     the neutrino event rate for $0 \lesssim T_{\rm pb} \lesssim $ 150 ms 
is approximately fitted by a linear function (as a function of postbounce time) with positive slope, whereas it can be fitted by a linear function with a negative slope plus 
 the SASI modulation thereafter. When we evaluate the correlation function
 $X(t,\Delta T)$ in Eq. (\ref{eq:correlationfunction}), the large offset can be a 
hinderance for an appropriate evaluation of $X(t,\Delta T)$.
     We thus roughly remove the quasi-monotonically changing component, 
i.e., the offset, in a simple way as $A_\nu(t)\rightarrow A_\nu(t)-
     \left(A_\nu(t+\tau/2)+A_\nu(t-\tau/2)\right)/2$. Here $\tau$ is a time window and we usually use $\tau=60$ ms.
 }.
 
 We then evaluate the correlation between 
the GW (blue line) and neutrino (red line) signal
by calculating the following normalized correlation function $X(t,\Delta T)$
\begin{eqnarray}
\label{eq:correlationfunction}
X(t,\Delta T)=\frac{\int d\tau H(t-\tau)A_\nu(\tau+\Delta T)A_{\rm GW}(\tau)}
{\sqrt{\int d\tau H(t-\tau)(A_\nu(\tau+\Delta T))^2}
\sqrt{\int d\tau H(t-\tau)(A_{\rm GW}(\tau))^2}},
\end{eqnarray}
 where $t$ is the postbounce time and $H(t-\tau)$ is the Hann window with the 
window size of $|t-\tau|\le10$ ms. 
$A_\nu(t)$ and $A_{\rm GW}(t)$ is the neutrino event rate without the DC component
 and the GW amplitude, respectively.
$\Delta T$ [ms] represents the time delay between the neutrino and GW signal
and we take $0\le \Delta T\le24$ ms with time interval of 4 ms.
In the middle panel, we plot $X(t,\Delta T)$ for all $\Delta T$ in different 
colors as shown in the upper part of the panel. 
The bottom panel shows $\Delta T_{\rm max}$, 
which gives the maximum $X(t,\Delta T)$.

From the middle panels of Figure \ref{f8}, we find a clear increment 
in $|X(t,\Delta T)|$ at $T_{\rm pb}\sim150$ ms for both GW polarized modes.
At this point of time, the SASI activity becomes strongest (see Fig.3 in \cite{KurodaT16ApJL}).
If we look at panel (b+), $|X(t,\Delta T)|$ with larger $\Delta T$ increases 
faster. $X(t,\Delta T=24 {\rm ms})$ increases fast with positive value and 
then $X(t,\Delta T=20 {\rm ms})$ comes next, but with negative value. 
Afterward, $X(t,\Delta T=16 {\rm ms})$, $X(t,\Delta T=12 {\rm ms})$, $\cdots$,
 follow with the same manner. Completely an opposite trend can be seen in the 
 $\times$ mode of the polarization (panel (b$\times$)),
i.e., $X(t,\Delta T=24 {\rm ms})$, $X(t,\Delta T=16 {\rm ms})$, $\cdots$, show 
negative values and $X(t,\Delta T=20 {\rm ms})$, $X(t,\Delta T=12 {\rm ms})$, 
$\cdots$, show positive values. 
This can be explained by the correlation frequency $F\sim120$ Hz 
(see top left panel in Fig. \ref{f5} and left one in Fig. \ref{f6}).
The corresponding time period $\sim8$ ms of $F\sim120$ Hz leads to a cycle of 
negative and positive correlations
if we shift neutrino count event with half of its value, i.e., $\sim4$ ms.
Furthermore the opposite trend between (b+) and (b$\times$) can be understood by 
the phase shift with half period between the plus and cross mode of GWs, 
since the leading term of the PNS deformation is the quadrupole 
($l=2$) mode \citep{KurodaT16ApJL}.
From panels (c+/$\times$), $\Delta T_{\rm max}$ with $\sim 18$ 
ms appears first in both polarization modes.
It means that there is a time delay of GWs from neutrinos as $\Delta T\sim18$ ms.
Remarkably this value is consistent with our previous rough measurement for the 
accretion timescale of a few 10 ms. Note that we have also done the same analysis 
for the rest of our models and found no significant correlation.
As a reference, Figure \ref{f9} is shown for S11.2(SFHx) where
 there is no significant correlation between the GW and neutrino signals 
for this convection-dominated model.

\section{Summary and Discussion}\label{sec6}
We have presented results from our 3D-GR core-collapse simulations 
with approximate neutrino transport for three non-rotating progenitors 
(11.2, 15, and 40 $M_{\odot}$) using three different EOSs. 
 Among the five computed models, the SASI activity was 
only unseen for an $11.2 M_{\odot}$ star. We have found that the 
combination of progenitor's higher compactness at bounce and the use of 
  softer EOS leads to the stronger SASI activity. Our 3D-GR models have confirmed previous predications that
 the SASI produces characteristic time modulations 
both in the neutrino and GW signals. Among the computed models, a
 15.0 $M_{\odot}$ model using SFHx EOS exhibited the most violent SASI motion, 
 where the SASI-induced modulation in both 
GWs and neutrinos were most clearly observed. The typical modulation frequency
 is in the range of $\sim100$-200 Hz, which is consistent with the oscillation
 period of the SASI motion.
 By performing a correlation analysis
 between the SASI-induced neutrino and GW signatures, we have found that the correlation becomes highest when we take into account the 
  time-delay effect due to the advection of material from 
the neutrino sphere to the PNS core surface. Our results suggest that 
the correlation of the neutrino and GW signals, if detected, could provide a 
 new signature of the vigorous SASI activity in the supernova core, which
 can be barely seen (like for the 11.2 $M_{\odot}$ model) if
 neutrino-convection dominates over the SASI.

In order to enhance predicative power of the neutrino and GW signals
 in this work, we need to at least update our M1 scheme from 
 gray to multi-energy transport as in \citet{KurodaT16,roberts16}.
 Inclusion of detailed neutrino opacities is also mendatory 
(e.g., \citet{Buras06a,lentz12a,gabriel12,tobias16,burrows16,horowitz17,roberts17,bollig17}). Impacts of rotation and magnetic fields 
\citep{moesta14,takiwaki16,martin17} on the correlation between 
the GW and neutrino signals \citep{Ott12,yokozawa} should be 
 also revisited with 3D-GR models including more sophisticated
 neutrino transport scheme with elaborate neutrino opacities.

In order to clarify whether we can or cannot detect the SASI-induced modulation
 in the GW and neutrino signals, we primarily need to perform a GW signal 
 reconstruction study (e.g., \citet{Hayama15,Powell16,Gossan16}) 
 using non-stationary and non-Gaussian noise \citep{Powell16,Powell17}.
 This is the most urgent task that we have to investigate as a sequel of this work.
For a Galactic event, we apparently need third-generation
 detectors for observing the SASI-modulated GW signals (e.g., \citet{Andresen17}),
 whereas the neutrino signals could be surely detected by IceCube 
and Super-K \citep{Tamborra13}.
 The neutrino burst can be used to determine the core bounce time
 \citep{Halzen09}, which  raises significantly the detection efficiency of the GWs 
\citep[e.g.,][]{Gossan16,nakamura16}.
Our current study extends the horizon of previous prediction such as,
 when we would succeed the simultaneous detection of neutrino and GW 
signals from future nearby CCSN event, we could infer the supernova 
 triggering dynamics (e.g., the SASI) from the following 
specific features (1) the low frequency ($F\sim100$ Hz) modulation 
 in both GW and neutrino signal and (2) a few 10 
ms time delay of the SASI-modulated GW signal from the SASI-modulated
 neutrino event rate. 
We finally note that the non-detection of the correlation 
could be hypothetically used as a 
measure to constrain the nuclear EOSs. From our limited number of the EOS used 
in this work, one cannot obtain any quantitative conclusion. Recently, 
a number of nuclear EOS is available (see \cite{Oertel17} for a review). Using such
 rich variety of the EOSs, one could in principle do this, but only if one could 
afford enough computational time to make the many 3D CCSN runs doable. 

 

\acknowledgements{
TK was supported by the European Research Council (ERC; FP7) under ERC Advanced Grant Agreement N$^\circ$ 321263 - FISH and ERC StG EUROPIUM-677912.
Numerical computations were carried out on Cray XC30 at Center for Computational Astrophysics, National Astronomical Observatory of Japan. KK was thankful to stimulating discussions with E. M\"uller, H.T. Janka, 
and T. Foglizzo.
This study was supported by JSPS KAKENHI Grant Number (JP15H00789, JP15H01039,
JP15KK0173, JP17H01130, JP17H05206, JP17K14306, and JP17H06364) and
JICFuS as a priority issue to be tackled by using Post `K' Computer.
}

%



\bibliographystyle{apj}
\bibliography{mybib}

\end{document}